\documentclass[longauth]{aa}  

\usepackage{graphicx}
\usepackage{txfonts}
\usepackage{hyperref}
\usepackage{booktabs}
\usepackage{orcidlink}
\usepackage{multirow}
\usepackage{tabularray}
\usepackage{float}
\usepackage{afterpage}
\usepackage{placeins}

\begin{document}

   \title{Combining the second data release of the European Pulsar Timing Array with low-frequency pulsar data}

   \author{F.~Iraci\orcidlink{0009-0001-0068-4727}\inst{\ref{UniCA},\ref{inaf-oac}}\fnmsep\thanks{\email{francesco.iraci@inaf.it}}
          \and
          A.~Chalumeau\orcidlink{0000-0003-2111-1001}\inst{\ref{astron}}\fnmsep\thanks{\email{chalumeau@astron.nl}}
          \and
          C.~Tiburzi\orcidlink{0000-0001-66-51-4811}\inst{\ref{inaf-oac}}
          \and
          J.~P.~W.~Verbiest\orcidlink{0000-0002-4088-896X}\inst{\ref{fsi}}
          \and
          A.~Possenti\orcidlink{0000-0001-5902-3731}\inst{\ref{inaf-oac}}
          \and
          S.~C.~Susarla\orcidlink{0000-0003-4332-8201}\inst{\ref{uni-galway}}
          \and
          M.~A.~Krishnakumar\orcidlink{0000-0003-4528-2745}\inst{\ref{ncra}}
          \and
          G.~M.~Shaifullah\orcidlink{0000-0002-8452-4834}\inst{\ref{uni-MiBicocca}\ref{infn-uniMiB}}
          \and
          J.~Antoniadis\orcidlink{0000-0003-4453-3776}\inst{\ref{forth-greece}\ref{maxplanck-bonn}}
          \and
          M.~Bagchi\orcidlink{0000-0001-8640-8186}\inst{\ref{hbni-india}\ref{mathscienceinst-india}}
          \and
          C.~Bassa\orcidlink{0000-0002-1429-9010}\inst{\ref{astron}}
          \and
          R.~N.~Caballero\orcidlink{0000-0001-9084-9427}\inst{\ref{kavli-beijing}}
          \and
          B.~Cecconi\orcidlink{0000-0001-7915-5571}\inst{\ref{LIRA-france}\ref{ORN-nancay}}
          \and
          S.~Chen\orcidlink{0000-0002-3118-5963}\inst{\ref{shanghai-china}\ref{lab-radioastro-Beijing}}
          \and
          S.~Chowdhury\inst{\ref{mathscienceinst-india}}
          \and
          B.~Ciardi\inst{\ref{maxplank-Garching}}
          \and
          I.~Cognard\orcidlink{0000-0002-1775-9692}\inst{\ref{LPC2E}}
          \and
          S.~Corbel\orcidlink{0000-0001-5538-5831}\inst{\ref{uni-cite-saclay}}
          \and
          S.~Desai\inst{\ref{IIT-KandiIndia}}
          \and
          D.~Deb\orcidlink{0000-0003-4067-5283}\inst{\ref{mathscienceinst-india}}
          \and
          J.~Girard\orcidlink{0000-0003-0432-403X}\inst{\ref{LIRA-france}}
          \and
          A.~Golden\orcidlink{0000-0001-8208-4292}\inst{\ref{uni-galway}}
          \and
          J-M.~Grie{\ss}meier\orcidlink{0000-0003-3362-7996}\inst{\ref{LPC2E}\ref{ORN-nancay}}
          \and
          L.~Guillemot\orcidlink{0000-0002-9049-8716}\inst{\ref{LPC2E}\ref{ORN-nancay}}
          \and
          M.~Hoeft\inst{\ref{tautenburg}}
          \and
          H.~Hu\orcidlink{0000-0002-3407-8071}\inst{\ref{maxplanck-bonn}}
          \and
          F.~Jankowski\orcidlink{0000-0002-6658-2811}\inst{\ref{LPC2E}}
          \and
          G.~Janssen\orcidlink{0000-0003-3068-3677}\inst{\ref{astron}\ref{uni-Nijmegen NL}}
          \and
          B.~C.~Joshi\orcidlink{0000-0002-0863-7781}\inst{\ref{ncra}\ref{IITR-india}}
          \and
          S.~Kala\inst{\ref{mathscienceinst-india}}
          \and
          E.~Keane\orcidlink{0000-0002-4553-655X}\inst{\ref{trinityC-Dublin}}
          \and
          K.~Nobelson\orcidlink{0000-0003-2715-4504}\inst{\ref{uni-kunamoto japan}\ref{iroast-kunamoto_uni-japan}}
          \and
          A.~Konovalenko\orcidlink{0000-0003-1949-9625}\inst{\ref{inst-radioastro Ukraine}}
          \and
          I.~Kravtsov\orcidlink{0000-0002-2312-3025}\inst{\ref{inst-radioastro Ukraine}\ref{LPC2E}}
          \and
          M.~Kramer\inst{\ref{maxplanck-bonn}}
          \and
          K.~Liu\orcidlink{0000-0002-2953-7376}\inst{\ref{shanghai-china}\ref{lab-radioastro-Beijing}}
          \and
          A.~Parthasarathy\orcidlink{0000-0002-4140-5616}\inst{\ref{astron}\ref{uni-amsterdam}\ref{maxplanck-bonn}}
          \and
          P.~Rana\orcidlink{0000-0001-6184-5195}\inst{\ref{uni-capetown-Astronomy dep}\ref{depAstro-mumbai}}
          \and
          D.~Schwarz\orcidlink{0000-0003-2413-0881}\inst{\ref{uni-bielefeld}}
          \and
          J.~Singha\orcidlink{0000-0002-1636-9414}\inst{\ref{uni-capetown}}
          \and
          A.~Srivastava\orcidlink{0000-0003-3531-7887}\inst{\ref{uni-GLA-india}\ref{IIT-KandiIndia}}
          \and
          K.~Takahashi\orcidlink{0000-0002-3034-5769}\inst{\ref{uni-kunamoto japan}\ref{iroast-kunamoto_uni-japan}}
          \and
          P.~Tarafdar\inst{\ref{inaf-oac}}
          \and
          G.~Theureau\inst{\ref{LPC2E}\ref{ORN-nancay}}
          \and
          O.~Ulyanov\orcidlink{0000-0003-0934-0952}\inst{\ref{inst-radioastro Ukraine}}
          \and
          C.~Vocks\orcidlink{0000-0001-8583-8619}\inst{\ref{inst-potsdam}}
          \and
          J.~Wang\orcidlink{0000-0003-1933-6498}\inst{\ref{uni-bochum}}
          \and
          V.~Zakharenko\orcidlink{0000-0001-9977-824X}\inst{\ref{inst-radioastro Ukraine}}
          \and
          P.~Zarka\orcidlink{0000-0003-1672-9878}\inst{\ref{LIRA-france}}
          }

   \institute{Dipartimento di Fisica, Università di Cagliari, Cittadella Universitaria, I-09042 Monserrato (CA), Italy \label{UniCA}
   \and
   INAF - Osservatorio Astronomico di Cagliari, via della Scienza 5, 09047 Selargius (CA), Italy\label{inaf-oac}
   \and
    ASTRON, Netherlands Institute for Radio Astronomy, Oude Hoogeveensedijk 4, 7991 PD, Dwingeloo, The Netherlands\label{astron}
    \and
    Florida Space Institute, University of Central Florida, 12354 Research Parkway, Orlando, FL 32826, USA\label{fsi}
    \and
    Physics, School of Natural Sciences, Ollscoil na Gaillimhe --- University of Galway, University Road, Galway, H91 TK33, Ireland\label{uni-galway}
    \and
    National Centre for Radio Astrophysics, Pune University Campus, Pune 411007, India \label{ncra}
    \and
    Dipartimento di Fisica ``G. Occhialini'', Università degli Studi di Milano-Bicocca, Piazza della Scienza 3, I-20126 Milano, Italy\label{uni-MiBicocca}
    \and
    INFN, Sezione di Milano-Bicocca, Piazza della Scienza 3, I-20126 Milano, Italy\label{infn-uniMiB}
    \and
    Institute of Astrophysics, Foundation for Research \& Technology -- Hellas (FORTH), GR-70013 Heraklion, Greece\label{forth-greece}
    \and
    Max-Planck-Institut f\"{u}r Radioastronomie, Auf dem H\"{u}gel 69, DE-53121 Bonn, Germany\label{maxplanck-bonn}
    \and
    The Institute of Mathematical Sciences, C. I. T. Campus, Taramani, Chennai 600113, India \label{mathscienceinst-india}
    \and
    School of Physics, Trinity College Dublin, College Green, Dublin 2, D02 PN40, Ireland\label{trinityC-Dublin}
    \and
    Homi Bhabha National Institute, Training School Complex, Anushakti Nagar, Mumbai 400094, India \label{hbni-india}
    \and
    Max Planck Institute for Astrophysics, Karl-Schwarzschild-Str 1, D-85741 Garching, Germany \label{maxplank-Garching}
    \and
    Kavli Institute for Astronomy and Astrophysics, Peking University, Beijing 100871, China \label{kavli-beijing}
    \and
    LIRA, Observatoire de Paris, CNRS, PSL, Sorbonne U., U. Paris Cit\'e, Meudon, France\label{LIRA-france}
    \and
    ORN, Observatoire de Paris, CNRS, PSL, U. Orl\'eans, Nan\c{c}ay, France \label{ORN-nancay}
    \and
    Shanghai Astronomical Observatory, Chinese Academy of Sciences, Shanghai 200030, P. R. China\label{shanghai-china}
    \and
    Key Laboratory of Radio Astronomy and Technology, Chinese Academy of Sciences, Beijing 100101, P. R. China\label{lab-radioastro-Beijing}
    \and
    LPC2E, OSUC, Univ Orleans, CNRS, CNES, Observatoire de Paris, F-45071 Orleans, France\label{LPC2E}
    \and
     Université Paris Cité and Université Paris Saclay, CEA, CNRS, AIM, 91190 Gif-sur-Yvette, France \label{uni-cite-saclay}
     \and
     Th\"{u}ringer Landessternwarte, Sternwarte 5, 07778 Tautenburg, Germany\label{tautenburg}
     \and
     Department of Physics, IIT Hyderabad, Kandi Telangana 502284, India \label{IIT-KandiIndia}
     \and
     Department of Astrophysics/IMAPP, Radboud University Nijmegen, P.O. Box 9010, 6500 GL Nijmegen, The Netherlands\label{uni-Nijmegen NL}
     \and
     Indian Institute of Technology, Roorkee 247667 India \label{IITR-india}
     \and
     Faculty of Advanced Science and Technology, Kumamoto University, 2-39-1 Kurokami, Kumamoto 860-8555, Japan\label{uni-kunamoto japan}
     \and
     Institute of Radio Astronomy of the National Academy of Sciences of Ukraine, Mystetstv St. 4, Kharkiv, Ukraine 61002\label{inst-radioastro Ukraine}
     \and
     High Energy Physics, Cosmology \& Astrophysics Theory (HEPCAT) Group, Department of Mathematics and Applied Mathematics, University of Cape Town, Cape Town 7700, South Africa\label{uni-capetown}
     \and
     Department of Astronomy, University of Cape Town, Cape Town 7700, South Africa\label{uni-capetown-Astronomy dep}
     \and
     Department of Astronomy and Astrophysics,  Tata Institute of Fundamental Research, Colaba, Mumbai 400005\label{depAstro-mumbai}
     \and
     Anton Pannekoek Institute for Astronomy, University of Amsterdam, Science Park 904, 1098 XH Amsterdam, The Netherlands \label{uni-amsterdam}
     \and
     Fakult\"at f\"ur Physik, Universit\"at Bielefeld, Postfach 100131, 33501 Bielefeld, Germany\label{uni-bielefeld}
     \and
     Department of Physics, GLA University, Mathura 281406, India\label{uni-GLA-india}
     \and
     International Research Organization for Advanced Science and Technology, Kumamoto University, 2-39-1 Kurokami, Kumamoto 860-8555, Japan\label{iroast-kunamoto_uni-japan}
     \and
     Leibniz-Institut f\"ur Astrophysik Potsdam (AIP), An der Sternwarte 16, 14482 Potsdam, Germany\label{inst-potsdam}
     \and
     Ruhr-Universit\"at Bochum, Fakult\"at f\"ur Physik und Astronomie, Astronomisches Institut (AIRUB), 44801 Bochum, Germany\label{uni-bochum}
    }
   \date{Received XX; accepted XX}
  \abstract 
  {Low radio frequency data are highly valuable for enhancing the sensitivity of pulsar timing arrays (PTAs) to propagation effects, such as dispersion measure (DM) variations. 
  These low-frequency observations are particularly sensitive to DM fluctuations and can therefore significantly improve noise characterization in PTA datasets, which is essential for detecting the stochastic gravitational wave background (GWB).}
  {For this work we incorporated for the first time low-frequency observations from LOFAR ($100-200\,\mathrm{MHz}$) and NenuFAR ($30-90\,\mathrm{MHz}$) into a PTA context by combining them with the most recent data release from the European and Indian PTAs (in particular, with the subsample labeled \texttt{DR2new+}, which includes only data from the new backends).
  This new combined dataset, labeled \texttt{DR2low}, consists of 12 pulsars observed over a time span of $\sim$11 years, with radio frequencies spanning the range $30-2500\,\mathrm{MHz}$.
  The expanded frequency coverage of \texttt{DR2low} enables us to update and refine the noise models of \texttt{DR2new+}, and this is crucial in order to increase the PTA sensitivity when searching for the stochastic gravitational wave background, which is the primary goal of PTA observations.
  This work is a milestone in the integration of low-frequency data into the upcoming third data release of the International PTA, which is posed to achieve the 5$\sigma$ detection of the GWB.
  }
  {We used the pulsar timing software packages \textsc{Libstempo} and \textsc{Enterprise} to perform a noise analysis of \texttt{DR2low}.
  At first, we applied a \texttt{standard noise model} including red noise (RN) and time-variable dispersion measure (DMv) as  power laws, with Fourier components up to 30 and 100 frequencies, respectively.
  Next, we performed a fully Bayesian model selection to identify the favored noise model for each pulsar and compute the Bayes factors across all combinations of RN, DMv, and a noise term with a chromatic index of 4 (CN$_4$).
  Finally, we carried out a detailed analysis on the choice of the chromatic index for CN$_4$ and the contribution of the solar wind.}
  {The comparison between \texttt{DR2low} and \texttt{DR2new+} using the \texttt{standard noise model} highlights the benefits of including low-frequency data. 
  In particular, the additional frequency coverage improves the constraints on the DM variations and helps  disentangle the DM and RN noise components in most pulsars.
  Through a Bayesian model selection, we found that the RN is required in the final model for 10 out of 12 pulsars, compared to only 5 in the \texttt{DR2new+} dataset.
  The improved sensitivity to plasma effects provided by \texttt{DR2low} also favors the identification of significant CN$_4$ in eight pulsars, while none showed such evidence in \texttt{DR2new+}.
  The chromatic index of this process is consistent with four of the five pulsars, while two (PSRs J0030+0451 and J1022+1001) show significant deviations from such a value. 
  We attribute this discrepancy to unmodeled contributions from the solar wind, especially because of the high DM sensitivity of LOFAR and NenuFAR and the high observing cadence provided by these datasets near solar conjunction.
  A dedicated analysis confirms that the current solar wind model fails to fully capture the observed delay, and residual power is absorbed into the DM component of the model.
  }
   {}

   \titlerunning{EPTA DR2low}
   \authorrunning{F.~Iraci et al.}
   \maketitle

\section{Introduction}\label{sec:intro}
Pulsar timing array (PTA) experiments exploit the clock-like precision of the periodic broadband radio emission received from pulsars to detect low-frequency (on the order of the nHz order) gravitational waves (GWs). 
Unpredicted fluctuations in such a precise series of radiation pulses reflect any kind of perturbation that the pulsar signal experiences, ranging from irregularities in the pulsar’s rotation to the influence of Galactic plasma and GWs.
A stochastic and isotropic gravitational wave background (GWB), generated by a population of supermassive black hole binaries (SMBHBs), introduces a temporally correlated noise in the time-of-arrival (ToA) fluctuations—also known as timing residuals—relative to the deterministic timing model \citep{sesana2013}.
In the most simple case of circular SMBHBs, this noise has a red power spectrum, characterized by a steep spectral index of -13/3 \citep{p01}.
Moreover, these GW-induced fluctuations follow a spatial correlation among pulsars, referred to as the Hellings \& Downs curve \citep{hd83}.
To date, eight PTA collaborations exist in the world: the European PTA \citep[EPTA][]{epta_dr1_timing}, the North-American Nanohertz Observatory for Gravitational waves \citep[NANOGrav][]{ransom2019}, the Parkes PTA \citep[][]{manchester2013}, the Indian PTA \citep[InPTA][]{bcj2018J}, the Chinese PTA \citep[CPTA][]{lee2016}, the MeerKAT PTA \citep[MeerTIME][]{miles2023}, the African Pulsar Timing (APT\footnote{\url{https://africanpulsartiming.github.io}}) consortium, and the International PTA \citep[IPTA][]{vlh2016}. 
In 2023, five PTA collaborations presented results supporting the very first evidence for the presence of a GWB in PTA data \citep{epta23_gwb,ng23_gwb,ppta23_gwb,cpta23_gwb}, followed a year later by the sixth collaboration \citep{mpta24_gwb}. 
Although none of these studies proved the GWB presence with a significance higher than the required 5$\sigma$ threshold \citep{Allen2023}, the results will be improved by the IPTA (formed by EPTA, NANOGrav, PPTA, InPTA, MeerTIME, and APT), which will combine the data from all of the individual PTAs. 
The EPTA will not only contribute with its second data release (described in \cite{epta23_timing} and \cite{epta23_noise}), but also by enhancing the IPTA dataset. 
The combination of PTA datasets within the IPTA framework improves the overall electromagnetic frequency-domain coverage, increases the sky coverage by including more pulsars, and benefits from a higher observing cadence. 
However, certain noise contributions remain challenging to model, particularly due to the limited availability of ultra-low-frequency observations spanning long time baselines.

The delay experienced by a radio wave as it propagates through the ionized interstellar medium (IISM) is inversely proportional to its frequency. 
Consequently, low-frequency pulsar data are especially sensitive to plasma-induced effects within the IISM through which the signal travels.
The most relevant propagation effects for PTAs are dispersion and, albeit less impacting, scattering. 
Dispersion indicates a phenomenon for which the group velocity $v_g$ of the propagating light changes, depending on its radio frequency, due to the variation in the refractive index $\mu$. In particular,
\begin{equation}
    \mu \approx \sqrt{1 - \left( \frac{f_p}{\nu} \right )^2},
\end{equation}
where $f_p$ is the plasma frequency and $\nu$ the observing (radio) frequency. 
The frequency-dependent delay $\delta t$ can be expressed as
\begin{equation}
    \delta t \approx \mathcal{D} \frac{\mathrm{DM}}{\nu^2},
\end{equation}
where $\mathcal{D}$ is the dispersion constant, and DM the dispersion measure (DM),  expressed as
\begin{equation}
    DM = \int_{\mathrm{LoS}} n_e dl ,
\end{equation}
where LoS is the line of sight and $n_e$ the electron density.

Scattering implies that the inhomogeneity distribution in the IISM generates refraction of the incoming wavefronts, so that their propagation path changes. 
Hence, a certain fraction of rays reaches the observer with a delay relative to those that are not deviated and follow the LoS, causing the pulse profile to broaden and develop an exponential tail. 
The frequency-dependence of this tail is parameterized  by the scattering timescale $\tau$, which scales as
\begin{equation}
    \tau \propto \nu^{-\gamma_{s}},
\end{equation}
where $\nu$ is the observing frequency and $\gamma_s$ is the chromatic index.
The parameter $\gamma_s$ varies with the kind of structure that induces the scattering and with the turbulence of the IISM.
It is predicted to be 4.4 for an extended Kolmogorov medium, although measurements deviate somewhat from this value \citep{BCC04}. 

The parameters DM and $\tau$ (or even $\gamma_s$) can be time-dependent, as most pulsars are fast-moving objects due to the kick they receive during the supernova explosion \citep{Hansen1997, Hobbs2005, Kaplin2023}.  
This motion causes the pulsar signal to cross different regions of the IISM, leading primarily to variable dispersion  and, to a lesser degree, scattering, both of which can contribute significantly to red noise in PTAs.
Since both of these phenomena are inversely dependent on the radio frequency of the propagating radiation, low-frequency data are fundamental to disentangling them from the achromatic contribution to the noise budget, including the GWB signal \citep{lmh24,ict24}.
 
In this article we present the combination of the second data releases of the EPTA and InPTA (hereafter DR2) with the corresponding datasets from the LOw Frequency ARray \citep[LOFAR,][]{hwg13} and the New extension in Nan\c{c}ay upgrading LOFAR \citep[NenuFAR,][]{bgt21}, along with the associated noise analysis.
In Section~\ref{sec:data} we review the characteristics of DR2 and of the datasets collected using LOFAR and NenuFAR, and present their combination and basic timing properties in Section \ref{sec:combtiming}.
In Section~\ref{sec:noise} we outline the approach and tools used to characterize the noise in the combined dataset.
In Section~\ref{sec:basic} we present the results of the basic noise runs and compare them with those from EPTA DR2.
In Section~\ref{sec:advanced} we describe the more advanced model selection analysis, and highlight the differences with respect to the EPTA DR2 results.
In Section~\ref{sec:sw} we examine the impact of the solar wind on the noise analysis.
Finally, we draw our conclusions in Section \ref{sec:conclusion}.

%-----------------------------------------------------------------
\section{Data description}\label{sec:data}

\subsection{EPTA-InPTA DR2}\label{subsec: EPTA dataset}
The data combination and noise analysis for the EPTA/InPTA are presented in \cite{epta23_timing} and \cite{epta23_noise} respectively.
In this data release, EPTA and InPTA  combined data collected from seven main European and Indian radio telescopes, i.e., the Nançay Radio Telescope (NRT), the 100-m Telescope
at Effelsberg, the Lovell Telescope, the upgraded Giant Meterwave Radio 
Telescope (GMRT), the Westerbork Synthesis
Radio Telescope, the Sardinia Radio Telescope and ``Large European Array for Pulsars'' (LEAP), from mid-1990s to 2021. 
Instead of analyzing the 42 pulsars presented in the first European data release \citep{epta_dr1_timing}, the authors selected 25 pulsars only, following an approach to maximize the sensitivity to the GWB and continuous GWs with a minimum number of pulsars \citep{spf22}.
These pulsars and their characteristics are reported in Tables B1-B7 of \cite{epta23_timing}.
The data were combined using time offsets known as \texttt{JUMPs} with respect to the reference dataset, chosen to be the L-band NUPPI backend from the NRT. 
After that, initial timing solutions were obtained using the \textsc{TEMPO2} timing software \citep{hem06}, and refined using the \textsc{TEMPONEST} Bayesian toolkit \citep{lah14} to introduce white noise (WN) parameters, red noise (RN; e.g., originating from rotational instabilities), and chromatic noise models (i.e., long-period noise that has some dependence on the observing frequency).

By considering only the new EPTA backends—more precise and more uniformly distributed across the frequency bands—combined with the InPTA dataset, the so-called \texttt{DR2new+} dataset was constructed. 
This combination showed evidence for a GWB with a statistical significance exceeding $3.5\sigma$ \citep{epta23_gwb}.
The importance of radio-frequency coverage in the EPTA data was further investigated in \citep{ffs25}, which demonstrated that poor frequency coverage can significantly reduce the sensitivity of PTAs to a GWB signal.
As discussed in Section~\ref{sec:combtiming}, this work focuses exclusively on the \texttt{DR2new+} dataset.

\subsection{LOFAR}\label{subsec: LOFAR dataset}
LOFAR \citep{hwg13} is a European low-frequency interferometer operating from approximately $30$ to $240$~MHz. 
Its largest concentration of modules (called stations) is located in the Netherlands, which also hosts the LOFAR core (i.e., a subsection of the interferometer that can form its own tied-array beam), while other international stations are distributed in Germany (six), Poland (three), France, the UK, Ireland, Sweden and Latvia (one each).
Each station is an interferometer in itself, and is provided with a field of high-band antennae (HBA) operating from 100 to 240~MHz, and low-band antennae (LBA) operating below 100~MHz. 
Each international station can be detached from the largest interferometer and used as a stand-alone telescope.

Since 2013 a large pulsar monitoring campaign has been conducted with a main sub-band of the HBA fields (from about 100 to 200~MHz band) of the LOFAR core (LC) and the six German international stations (named DE601, DE602, DE603, DE604, DE605 and DE609) -- some details are given by \cite{kvH2016,bkk2016,dvt20, jdthesis_2022}.
More than 100 pulsars were observed with a monthly to bimonthly cadence with the core, and with a weekly cadence with the German stations, and similar observing programs started in many other countries hosting the international stations.
Among the observed sources, 12 millisecond pulsars (MSPs) are in common with the twenty-five presented in EPTA DR2.

To calculate the ToAs from the described LOFAR dataset, we followed the methodology outlined in \citet{dvt20} and \citet{tsb21}. 
This involved constructing a frequency-resolved, high signal-to-noise ratio (S/N) profile template using either a single bright observation or an average of several high S/N profiles.
We then cross-correlated each channel from every observation with the corresponding template profile at the that frequency, using the method described by \cite{taylor1992} but with the uncertainties calculated through a Monte-Carlo routine as recommended by \cite{vlh2016}.
Based on the work by \cite{dvt20}, we generally used eight frequency channels (and we therefore calculated 8 initial ToAs per observation per pulsar), but this can vary depending on the pulsar's S/N. 
The ToAs are then checked for outliers using a Huber regression-based algorithm described in \citep{tvs19}.

\subsection{NenuFAR}\label{subsec: NenuFAR dataset}
NenuFAR \citep{bgt21} is a low-frequency French interferometer that covers the 10–85~MHz range, featuring enhanced sensitivity and a significantly improved passband compared to the LBA field of any LOFAR station, including those in the core.
As part of the NenuFAR Pulsar Key Project, a pulsar monitoring program has been ongoing since 2019, following a bi-weekly to monthly cadence and targeting approximately 40 pulsars — including two MSPs also present in the EPTA DR2 list: PSRs J0030+0451 and J1022+1001.

To calculate the ToAs for the NenuFAR pulsars, we first constructed a frequency-resolved, high S/N template profile based on the DM-corrected average of all observations taken when the pulsar was located more than 45$^\circ$ from the Sun. 
This selection helps avoid contamination from solar dispersion effects, which are non-trivial to correct.
As with the LOFAR dataset, ToAs are then obtained by a) partial averaging in frequency of both the observations and the template in order to achieve eight frequency channels with reasonable S/N; b) channel-wise cross correlating the frequency resolved template with the individual observations; and c) cleaning of outliers using the aforementioned software based on the Huber-regression scheme.

\section{Data combination and basic timing}\label{sec:combtiming}

\begin{figure*}[ht!]
    \centering
    \includegraphics[width=\linewidth]{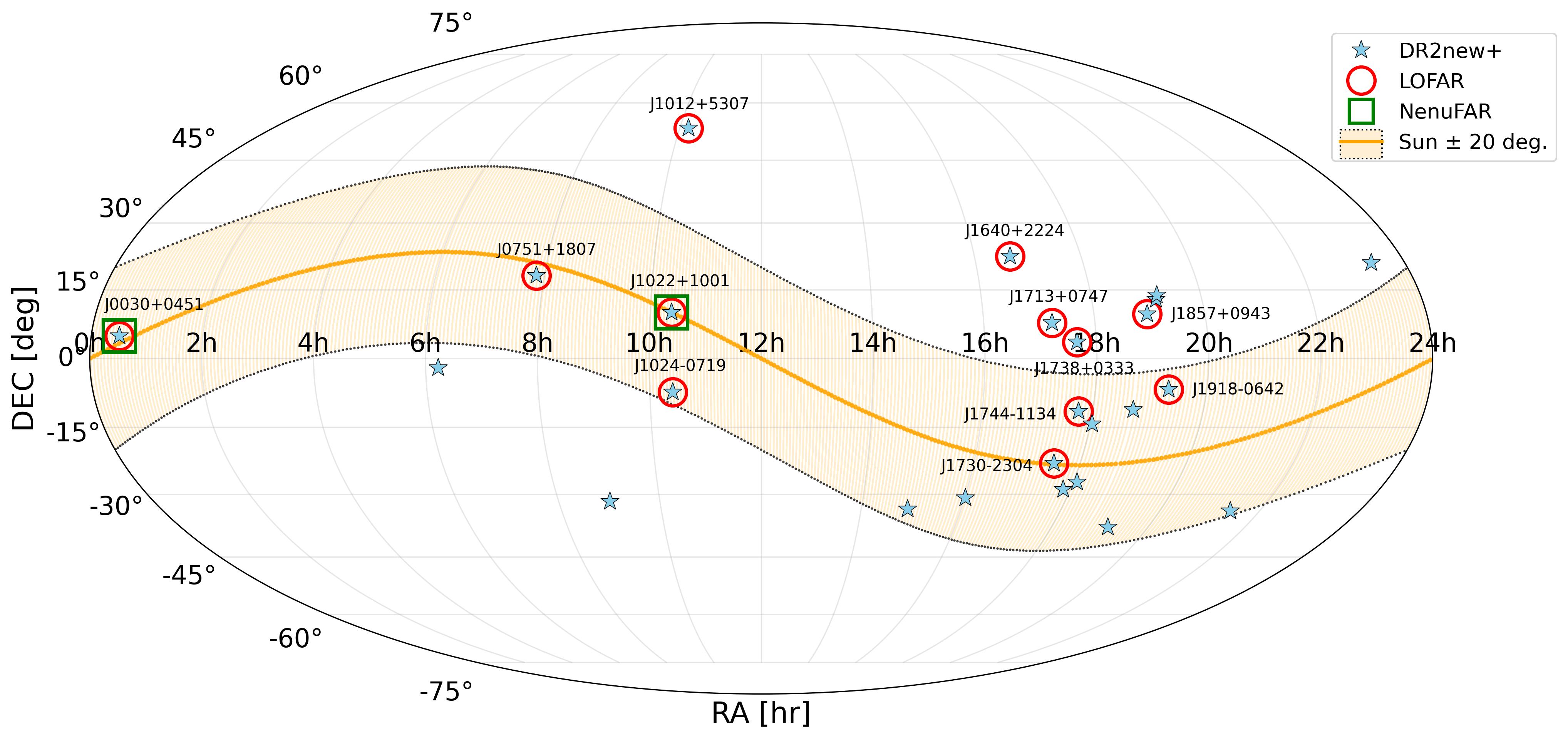}
    \caption{Mollweide sky projection of the 25 \texttt{DR2new+} pulsars (blue stars) in equatorial coordinates. 
    The red circles and green squares display sources observed by respectively LOFAR and NenuFAR. The names of the pulsars studied in this work are written next to the position. 
    The orange line (and area) are representing the Sun's position (± 20 degrees angle) along the year, denoting the influence zone of the solar wind.}
    \label{fig: Sky projection}
\end{figure*}

In the context of PTA studies, combining data from different observatories is essential to increase the overall data volume (number of observations), the frequency coverage and the time span, thereby improving sensitivity to the stochastic GWB \citep{sej13}. 
This section outlines the procedures used to combine the low radio frequency datasets of LOFAR and NenuFAR with \texttt{DR2new+} and describes the basic timing analysis performed to prepare the data for the noise analysis.
In Figure \ref{fig: Sky projection} we display the sky positions of the 25 MSPs from the \texttt{DR2new+}, emphasizing the names of the 12 sources studied in this work.
For the sake of clarity, the dataset composed of the combination of \texttt{DR2new+}, LOFAR, and NenuFAR is referred to as \texttt{DR2low} in the rest of the document. 

To address systematic time delays between observing systems, we fit for time offsets (\texttt{JUMPs}) for each observing station.
Specifically, we included one \texttt{JUMP} for each international LOFAR station, one for the LOFAR core, and one for NenuFAR when present.
Another important aspect to address when combining frequency-resolved data (one ToA per frequency channel, timed against the relevant frequency channels of a frequency-resolved template) is the difference between the DMs that were used to de-disperse the template profiles.
More precisely put, when timing frequency-resolved data against a frequency-resolved template, the ToAs of each frequency channel are measured relative to a phase reference that is defined by the relevant frequency channel of the template.
Through dedispersion of the template profile, the relative phases of the different channels are changed with respect to one another, introducing a frequency dependence of the ToA's phase reference.
This frequency-dependent phase reference is different for each template used and is defined by the combination of the DM used to dedisperse the template and the actual interstellar DM affecting the data that were used to construct the template.
Consequently, this effect is fully covariant (and indistinguishable from) a different DM affecting the datasets derived from a given template profile.
In order to account for this effect, we therefore applied a constant DM offset to all but one of the frequency-resolved datasets.
Specifically, in our analysis, frequency-resolved ToAs were available for the LOFAR, NenuFAR, and InPTA data, the latter of which was used as a reference for the other two to compute the DM offset.

Once the datasets have been prepared, we started from the EPTA/InPTA timing model and included the additional \texttt{JUMPs}.
After fitting the combined dataset, we observed that only a few key parameters exhibit significant changes—namely, the DM and its derivative (if present), the rotational period and its derivative, and the electron density at 1 AU to model the solar wind contribution.
All other timing model parameters remain consistent with their initial values, likely due to the larger ToA uncertainties in the low-frequency data.

In Figures \ref{fig:t_vs_freqs} and \ref{fig:t_vs_res}, we show the frequency coverage and timing residuals for each of the 12 combined pulsars. The ToAs, timing models, and low-frequency templates for all the pulsars are available on Zenodo\footnote{\url{https://doi.org/10.5281/zenodo.16794085}}.

\section{Noise analysis:  Methodology}\label{sec:noise}

\subsection{Framework and tools}
Having the combined dataset, we performed various iterations of noise analysis \citep{epta23_noise}, compared the results after including low radio frequency ($<300$~MHz) data, and assessed its improvement on constraining the models. 
The noise analyses are performed in a Bayesian framework, using the software \textsc{Libstempo}\footnote{https://github.com/vallis/libstempo} to extract data products from \textsc{TEMPO2}. 
The software \textsc{Enterprise} \citep{ent} and its complementary module \textsc{Enterprise\_Extensions} \citep{ent_ext} are used to model the noise components and build the likelihood and prior functions, applied to the MCMC sampler \textsc{PTMCMCSampler} \citep{ptmcmc}.

The remaining signals from the timing analysis described in Section \ref{sec:combtiming}, referred to as \texttt{timing residuals} $\delta t$, contain any effects impacting ToAs that are not included in the deterministic timing model. The analysis is performed by including the noise components in the multivariate Gaussian time-domain likelihood function, which thus assumes a Gaussian distribution of the noise-reduced timing residuals, as

\begin{equation}
\rm{log} \ L(\vec{\delta t}|\vec{\theta},\vec{\theta'}) = -\frac{1}{2} \left( \left[ \vec{\delta t} - \vec{y}(\vec{\theta'}) \right]^T \ C^{-1}(\vec{\theta}) \ \left[ \vec{\delta t} - y(\vec{\theta'})\right] \right) + \rm{cst.},
\end{equation}
where the timing residuals $\vec{\delta t}$ are reduced by any deterministic time-domain waveform $\vec{y}$ with parameters $\vec{\theta'}$, and where the covariance matrix is composed of the sum of all the covariance matrices from all the stochastic component $i$ with parameters $\theta_i$ such that $C(\vec{\theta}) = \sum_\lambda C^\lambda(\vec{\theta_\lambda})$.
The following subsection describes the noise components used in this work and shows how they are included in the likelihood function.

\subsection{Model components}
The noise components used in this work are similar to those from \cite{epta23_noise}, and allow a one-to-one comparison.
However, given the high sensitivity of low-frequency data to the solar wind dispersion, we add a refined model \citep{hsm22, Susarla2024} to include its constant and time-varying components that are fitted simultaneously with other noise components during the noise analysis. 
Table (\ref{tab:parprior}) describes all the components, their main properties and their parameter priors. \\

\subsubsection{White noise}

In the case of a perfect description of the ToAs from the timing model, the timing residuals divided by their uncertainties would follow a standard normal distribution. 
However, the measurement of the ToAs and their uncertainties (that estimate the so-called radiometer noise) are in practice subject to errors which have been found to be system-dependent. 
These errors are most commonly stochastic and time-uncorrelated, thus being "white noise" since they correspond to a flat power spectral density (PSD). 

In the case of the observing systems of \texttt{DR2new+}, we use the same approach as in \cite{epta23_noise}, with an Error FACtor (\texttt{EFAC}) applied to the provided ToA uncertainty $\sigma_{\rm{init, i}}$ and an Error QUADrature (\texttt{EQUAD}) term added to them in quadrature, for each observing system $\alpha$. 
\iffalse
As in \cite{epta23_noise}, this term is added in the likelihood covariance matrix $C$ as

\begin{equation}
    \label{eq:wn1}
    C^{\rm{WN,DR2new+}}_{\alpha} = \left( \texttt{EFAC}_{\alpha}^2 \times \sigma_{\rm{init, i}}^2 + \texttt{EQUAD}_{\alpha}^2 \right) \delta_{ij} \ \delta_{\alpha\beta},
\end{equation}
where the indices $i,j$ and $\alpha, \beta$ range over the ToAs and the observing systems, respectively. 
The symbol $\delta$ refers to the Kronecker delta in the whole document.
\fi
For the ToAs provided by LOFAR and NenuFAR, an Error CORRelated term (\texttt{ECORR}) is also added for each observing system to account for a white noise that is fully correlated among ToAs measured at the same epoch but at different radio frequencies, this type of error is typically associated with pulse phase jitter, but also any frequency-correlated systematics \citep{cs10,sc12,lk12} may affect \texttt{ECORR} value. 
The white noise terms are added to the likelihood covariance matrix $C$

\begin{equation}
    \label{eq:wn}
    \begin{split}
        &C^{\rm{WN}}_{\alpha} = \\
        &\begin{cases}
            \left( \texttt{EF}_{\alpha}^2 \times \sigma_{\rm{init, i}}^2 + \texttt{EQ}_{\alpha}^2 \right) \delta_{ij} \ \delta_{\alpha\beta} \text{, if $\alpha \notin$ \{LOFAR, NenuFAR\},}\\
            \left[ \left( \texttt{EF}_{\alpha}^2 \times \sigma_{\rm{init, i}}^2 + \texttt{EQ}_{\alpha}^2 \right) \ \delta_{ij} + \texttt{EC}_\alpha^2 \ \delta_{\rm{E}(i)\rm{E}(j)} \right] \ \delta_{\alpha\beta} \text{, else,}
        \end{cases}
    \end{split}
\end{equation}
where \texttt{EF}, \texttt{EQ}, and \texttt{EC} stand for \texttt{EFAC}, \texttt{EQUAD}, and \texttt{ECORR}, respectively, the indices $i,j$ and $\alpha, \beta$ range over the ToAs and the observing systems respectively and $\rm{E}(i)$ indexes the epoch of the ToA $i$. The symbol $\delta$ refers to the Kronecker delta, as in the following parts of this document.\\

\subsubsection{Time-correlated stochastic signals}

The time-correlated stochastic signals are modeled as Gaussian processes (GPs) following the approach described in \cite{vhv14}, using the covariance-function expansion to express the likelihood component of the noise $\lambda$ as
\begin{equation}
    \label{eq:GP}
    C^{\lambda,\rm{GP}}(t_i,t_j) = \sum_{\mu,\nu} \ F^\lambda_\mu(t) \ \Phi^\lambda_{\mu\nu} F^\lambda_\nu(t'),
\end{equation}
where $F$ is a set of basis functions, $\Phi$ is the covariance matrix of the Gaussian process in the weight-space view \citep{rw06}, and $\mu, \nu$ both range over the number of basis functions.

In this work, this approach is used to model both the achromatic red noise and time-varying chromatic signals arising from the heterogeneous nature of the IISM.
These include DM variations not captured by the DM1 and DM2 parameters, as well as an additional chromatic delay with a $\nu^{-4}$ frequency dependence, consistent with the scattering expected under the thin screen approximation with same-size inhomogeneities \citep{cs10}.\footnote{We note that the chromaticity of IISM-induced delays strongly depends on the specific properties of the electron distribution and may vary between pulsars. For instance, a $\nu^{-4.4}$ dependence is expected for a turbulent extended medium described by Kolmogorov theory, while refractive propagation effects could produce a steeper scaling, up to $\nu^{-6.4}$ \citep{sc17}.}
These three noise components (respectively labeled \texttt{RN}, \texttt{DMv} and \texttt{CN$_4$}) are modeled in the PSD domain from an incomplete Fourier basis approach where the basis functions are built as a set of sine and cosine functions for each PSD frequency $f$, and the Gaussian process covariance matrices are defined as

\begin{equation}
    \label{eq:GPPSD}
    \Phi^\lambda_{\mu\nu} = S^\lambda(f_\mu) \  \Delta f \ \delta_{\mu\nu},
\end{equation}
where $S(f_\mu)$ is the PSDs amplitude at frequency $f_\mu$ and $\Delta f$ is the spectral resolution, equal to the inverse of the total time span of the time series. 

Unless it is specified, the PSD of these three components are modeled as power laws:
\begin{equation}
    \label{eq:GPPSD}
    S^\lambda(f) = \frac{A_\lambda^2}{12\pi^2} \left( \frac{f}{f_{\rm{yr}}}\right)^{-\gamma_\lambda} f_{\rm{yr}}^{-3}.
\end{equation}
Here $f_{\rm{yr}}$ refers to the frequency of $1 \rm{yr}^{-1}$ and $A_\lambda$, $\gamma_\lambda$ are respectively the amplitude referred at $f_{\rm{yr}}$ and the spectral index of the noise component $\lambda$.

In our case, the different noise components are $\lambda = \{\texttt{RN}, \texttt{DMv}, \texttt{CN$_4$}\}$, which differ from their chromaticity (i.e., dependence on the observed radio-frequency $\nu$) that is added in the basis functions as $F^\lambda_\mu(t) \propto (\nu / \nu_{\rm{ref}})^{-\chi}$, where $\chi = 0, 2, 4$ for \texttt{RN}, \texttt{DMv} and \texttt{CN$_4$} respectively, and $\nu_{\rm{ref}}$ being a reference frequency set at $1.4$ GHz in this work.
While the chromatic index $\chi$ is fixed for these components, it is also possible to fit for this parameter simultaneously with other parameters and hyper-parameters of the model.

The timing model parameter uncertainties are also modeled as GPs also by using the covariance-function expansion (cf. Eq. \ref{eq:GP}), but they are analytically marginalized over. 
The basis functions are the timing model parameter design matrix and the Gaussian process covariance matrix is a diagonal matrix with very large values \citep{cbp22}.\\

\subsubsection{Solar wind}

The solar wind (SW) is an outflow of charged particles from the Sun which can significantly affect the DM especially for observations of pulsars near the ecliptic plane.
Properly modeling the impact of SW is crucial for precise timing analysis, especially in presence of very low-frequency data.
To account for this, we adopted a detailed SW model following \cite{Susarla2024} which contains a deterministic and a stochastic component.

We include a deterministic signal which account for the electron content $n_e$ as a function of the angular separation between the pulsar and the Sun sampling it from
\begin{equation}
    DM_{\mathrm{SW}} = n_e\frac{\rho}{r_e\sin{\rho}}\left[1~\mathrm{AU}\right]^2,
\end{equation}
where $\rho$ is the pulsar-Sun-observer angle, $n_e$ is the electron density at 1$\mathrm{AU}$ (denoted as NE\_SW in \texttt{TEMPO2}), and $r_e$ is the observatory-Sun distance \citep{tvs19, edwards2006}.
Additionally, we account for secular variations in the solar wind by incorporating a stochastic SW signature modeled with GPs. 
This approach, implemented as SWGP, is available in the chromatic noise module of \texttt{enterprise\_extensions}.
It uses a power spectral density described by the following power-law formula: 
\begin{equation} 
S_{\mathrm{SW}} = A_{\mathrm{SW}}^2\left( \frac{f}{\mathrm{yr}^{-1}} \right)^{-\gamma_{\mathrm{SW}}} \mathrm{yr}^3. 
\end{equation} 
Here $S_{\mathrm{SW}}$ is the power spectral density for SWGP, $A_{\mathrm{SW}}$ is the SW amplitude at a frequency of $1/1\mathrm{yr}$, $\gamma_{\mathrm{SW}}$ is the spectral index, and $f$ is the Fourier frequency.
We use the default value for the number of Fourier bins that goes up to 1.5yr$^{-1}$.

In this work, the SW model is specific to each pulsar and primarily depends on factors such as the ecliptic latitude, the ToA precision, and the observing cadence. 
In Table~\ref{tab:SWmodelling} we summarize how we modeled the SW for each pulsar, and provide a comparison with the model used in the EPTA DR2.

\iffalse
\begin{table*}[]
\centering
\begin{tabular}{|c|c|c|c|c|c|}
\toprule
\multirow{2}{*}{PSR name} & \multirow{2}{*}{ELAT (deg.)} & \multicolumn{2}{|c|}{SW modelling $-$ \texttt{DR2low}} & \multicolumn{2}{|c|}{SW modelling $-$ \texttt{DR2new+}}\\
& & Timing & Noise & Timing & Noise \\
\midrule
J1022+1001 & $-0.06$ & value with err (fitted) & \texttt{n\_earth} + \texttt{SWv} & $7.22 \pm err$ (fitted) & marg \\
J1730$-$2304 & $0.19$  & value with err (fitted) & \texttt{n\_earth} & 7.9 (fixed) & marg \\
J0030+0451 & $1.45$  & value with err (fitted) & \texttt{n\_earth} + \texttt{SWv} & 7.9 (fixed) & marg\\
J0751+1807 & $-2.81$ & 7.9 (fixed) & \texttt{n\_earth} + \texttt{SWv}  & 7.9 (fixed) & marg \\
J1744$-$1134 & $11.81$ & value with err (fitted) & \texttt{n\_earth} + \texttt{SWv} & 7.9 (fixed) & marg \\
J1918$-$0642 & $15.35$ & 4 (fixed) & marg  & 7.9 (fixed) & marg \\
J1024$-$0719 & $-16.04$ & 4 (fixed) & marg  & 7.9 (fixed) & marg \\
J1738+0333 & $26.88$ & 4 (fixed) & marg  & 7.9 (fixed) & marg \\
J1713+0747 & $30.70$ & 4 (fixed) & marg  & 7.9 (fixed) & marg \\
J1857+0943 & $32.32$ & 4 (fixed) & marg  & 7.9 (fixed) & marg \\
J1012+5307 & $38.76$ & 4 (fixed) & marg   & 7.9 (fixed) & marg \\
J1640+2224 & $44.06$ & 4 (fixed) & marg  & 7.9 (fixed) & marg \\
\bottomrule
\end{tabular}
\caption{Solar wind modeling in \texttt{DR2low}.}
\label{tab:SWmodelling}
\end{table*}
\fi

In the EPTA analysis, a fixed value of NE\_SW = 7.9 was used for all pulsars, except PSR~J1022+1001, and its error was marginalized over during the noise analysis, as for the other timing model parameters (the limitations of this approach have already been addressed by \citet{lpk25}).
For the combined dataset, the treatment of the SW contribution varies across pulsars depending on their ecliptic latitude (ELAT) and the availability of high-cadence observations. 
Pulsars with very low ELAT, such as PSRs J1022+1001 and J0030+0451, clearly exhibit solar wind signatures in their timing residuals. 
Although \citet{Susarla2024} recommend including only the deterministic component of the SW model for pulsars with ELAT below $3^\circ$, we also included the stochastic term, SWGP, in these two cases, due to the availability of NenuFAR ToAs, which were not part of the previous analysis.
We applied the same extended model (including SWGP) to PSR~J1744$-$1134, whose ELAT is above $3^\circ$, and to PSR J0751+1807, which was not included in \citet{Susarla2024}. 
The only exception is PSR~J1730$-$2304, for which we excluded the stochastic SW contribution due to the lack of high-cadence observations around solar conjunction epochs.
\\

\subsubsection{Exponential dip}

It is known a priori  for PSR~J1713+0747 that timing residuals contain multiple exponential dips \citep{Coles2015, Lam2018, Xu2021}: a brief radio-frequency dependent drop (where ToAs are measured before the time predicted by the timing model) followed by an exponential recovery for several weeks/months. 
Depending on the corresponding chromatic index $\chi^E$, the origin of these events are either considered to be related with a drop in the IISM content in the line-of-sight \citep[if $\chi^E = 2$]{Lam2018}, or related to magnetospheric effects if $\chi <2$ and if profile changes are observed at the time of the event. 
The latter scenario is favored for the only event present in the dataset used in this work that happened in 2016 ($\sim$ MJD $57510$), where \cite{grs20} found a significant profile change at this epoch and a chromatic index measured at $1.15^{+0.18}_{-0.19}$ with data from the Murriyang (Parkes) radio telescope, also being consistent with the value of $1.00^{+0.56}_{-0.49}$ measured with EPTA data \citep{cbp22}. 
As in \cite{epta23_noise}, we model this effect as a deterministic signal with a waveform defined as

\begin{equation}
    \label{eq:GPPSD}
    \begin{split}
    &y^E(A_E, \tau_E, t_{0,E}, \chi_E) =\\
    &\begin{cases}
        0 \text{, if \ $t < t_{0,E}$}\\
        A_E \ (\nu/1.4 \ \rm{GHz})^{-\chi_E} \ \rm{exp}\left( -\frac{t - t_{0,E}}{\tau_E}\right) \text{, if \ $t \geq t_{0,E}$}
    \end{cases},
    \end{split}
\end{equation}
where $A^E$, $\chi^E$, $t_0^E$, and $\tau^E$ are respectively the amplitude (in residual units), the chromatic index, the reference epoch (in MJD) and the relaxation time (in days) of the exponential dip modeled for a time $t$ and a radio-frequency $\nu$.
\\

\subsection{Parameter estimation and model selection}
In this work, we performed several analyses per pulsar, including parameter estimation and model selection methods, in order to (1) get information on the properties of the chosen noise components described below, (2) analyze the correlations between them, and (3) obtain the most favored combination among the noise components, which could be compared against the results published in \cite{epta23_noise}. 
This enabled us to understand the implications of including LOFAR and NenuFAR data with \texttt{DR2new+}.

For the sake of clarity, since we consider multiple noise models per pulsar, we describe the noise models from their Gaussian process components (apart from the timing model uncertainties), and the number of PSD frequencies used.
As an example, the noise model commonly adopted in PTA analyses includes marginalized timing model parameter uncertainties, white noise, an achromatic RN component, and DM variations. 
The RN and DM components are both modeled as a power-law processes, with 30 and 100 frequency bins respectively. 
This noise model will be referred to as \texttt{RN30+DMv100} throughout this document.

For the model selection, we applied a similar approach as the one described in \cite{epta23_noise}, where we defined a set of candidate noise models, fit for the number of frequency bins (i.e., $\nu$ in Eq. \ref{eq:GP}) used to model the power-law PSD GPs, and compute a Bayes factor among the different chosen models with their most favored number of frequency bins. 
The selected models are \texttt{WN} (white noise only), \texttt{WN+RN}, \texttt{WN+DMv}, \texttt{WN+RN+DMv}, \texttt{WN+DMv+CN$_4$}, \texttt{WN+RN+DMv+CN$_4$}, identical to the models in \cite{epta23_noise}.

In all runs, we included the SW model described in Table~\ref{tab:SWmodelling}, which consists of a deterministic component, \texttt{n\_earth}, and a stochastic term, SWGP for a specific set of pulsars.
Previous analyses based on LOFAR-only data indicated strong evidence in favor of this choice, with significantly improved model fit when the SWGP component is included. 
While this model may not fully capture all aspects of the solar wind signal-particularly around annual harmonics-it generally leads to a more accurate description of the data, with notably reduced white noise levels in several systems.

\section{Noise analysis - Results with standard models}\label{sec:basic}

We performed a parameter estimation analysis following the approach described in Section \ref{sec:noise} on two datasets, \texttt{DR2new+} and \texttt{DR2low}, by using \texttt{standard noise models}, composed of \texttt{TM}, \texttt{WN}, \texttt{RN30}, \texttt{DMv100} and the SW following the description above. 

\begin{figure}[ht!]
    \centering
    \includegraphics[width=.5\textwidth,height=20cm]{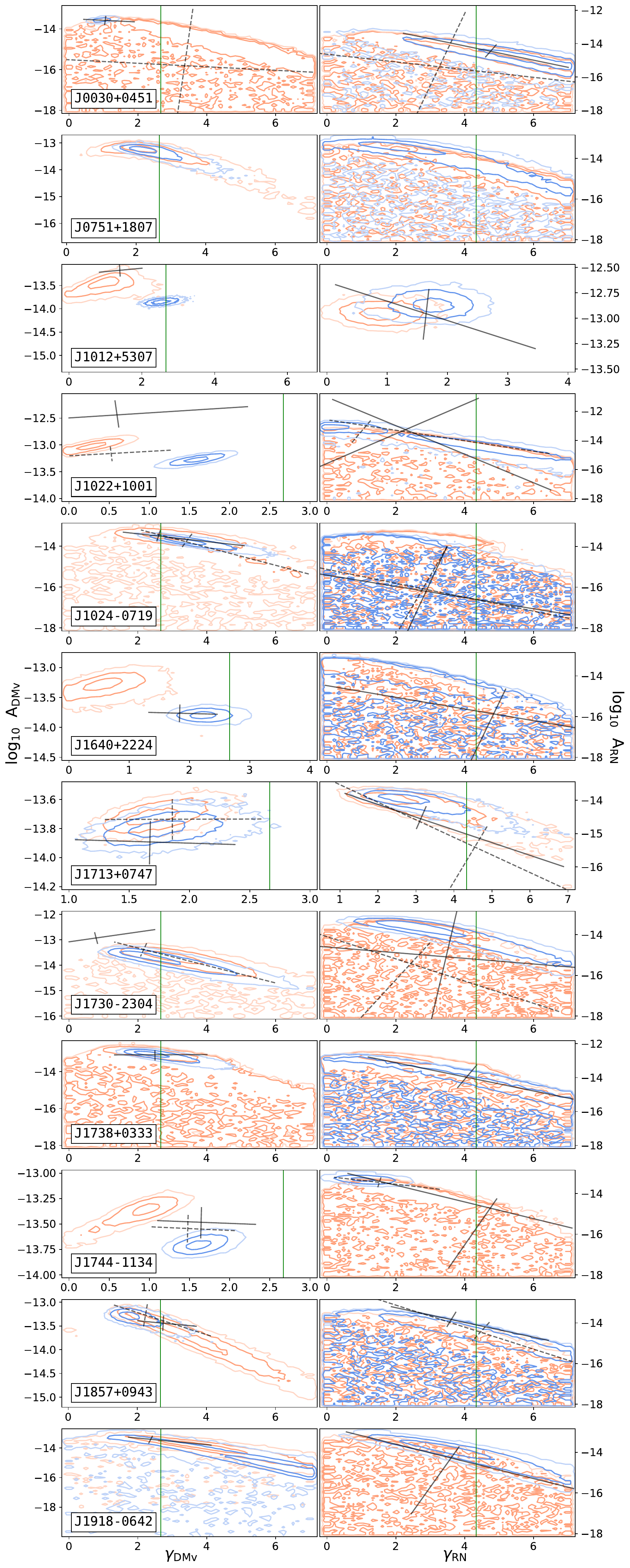}
    \caption{Spectral index vs. amplitude posterior distributions of the power-law PSD for DMv (left) and RN (right) obtained for 12 pulsars from \texttt{DR2low} (blue) and \texttt{DR2new+} (orange) using the standard noise models. The green vertical lines show the predicted spectral index from a Kolmogorov turbulence in the IISM for DMv (left), and from circular and GW-driven SMBHBs (right). The black crosses display the $99.7\%$ credible intervals of posterior distributions obtained with NANOGrav-15yr and PPTA DR3, displaying the principal directions.}
    \label{fig:corner_vanilla}
\end{figure}

In Figure~\ref{fig:corner_vanilla} we show the credible intervals (at confidence levels equivalent to 1, 2 and 3-$\sigma$ in a Gaussian distribution) of the noise parameter posterior distributions for \texttt{DR2new+} (orange) and \texttt{DR2low} (blue). For a better clarity, the color is lighter for the 3-$\sigma$ contour levels. 
The left-hand column presents the DM variations posterior distributions, with $\log_{10}A_{DM}$ on the y-axis and $\gamma_{DM}$ on the x-axis, while the right-hand column displays the corresponding parameters for the RN. 
For reference, we display the two principal directions\footnote{the principal directions are evaluated by projecting the distributions onto the eigenvectors, with asymmetric scaling factors corresponding to 3-$\sigma$ in a Gaussian distribution.} of the posteriors distributions for the same parameters obtained in \cite{3P+comparison2024} on the NANOGrav 15-year dataset \citep{ng15_timing} and PPTA DR3 \citep{PPTADR3_timing} with solid and dashed lines respectively.
The green vertical lines indicate the spectral index values expected from Kolmogorov turbulence (for the DM variations; left column), and the one predicted from a population of GW-driven supermassive black hole binaries with circular orbits (for the RN; right column).

\subsection{Observations for the DM variations}
\subsubsection{Comparing precision between DR2new+ and DR2low}

As expected, adding very low-frequency data from LOFAR and NenuFAR  provides stronger constraints of the \texttt{DMv} parameters (i.e., \texttt{DR2low} having narrower posterior distributions than \texttt{DR2new+}) for all the 12 pulsars, apart for PSRs J1022+1001, J1713+0747 and J1918$-$0642 for which they remain similar. 
In particular, we observed a very significant improvement for PSRs J0030+0451, J1024$-$0719, J1730$-$2304 and J1738+0333, which remained largely unconstrained in the \texttt{DR2new+} analysis.

\subsubsection{Consistency between DR2new+ and DR2low}

The new constraints remain fully consistent with \texttt{DR2new+} for 7 of the 12 pulsars: PSRs J0030+0451, J0751+1807, J1024$-$0719, J1730$-$2304, J1738+0333, J1857+0943 and J1918$-$0642, while a slight shift toward lower amplitudes is observed for PSR~J1713+0747, with a tension of only $\sim 0.77\sigma$ according to the description of a tension metric described in \cite{epta23_noise} based on \cite{rd21}. 
For PSR J1918$-$0642, we actually observed a slight increase of the wide posterior distribution at the high spectral index--low amplitudes in the parameter space.
However, the constraints are significantly different for PSRs J1012+5307, J1022+1001, J1640+2224, and J1744$-$1134, where the posterior distributions are shifted toward higher spectral indices (from about 1 to 2) and lower amplitudes (logarithmic values $\sim$0.3 to 0.5).
These discrepancies may arise from several factors. 
First, the \texttt{DR2new+} dataset mainly includes observations at high frequencies, which may limit its ability to disentangle chromatic (DMv) and achromatic (RN) noise components.
For PSR~J1640+2224, \texttt{DR2new+} contains only data at L and S bands, making it inherently unreliable for DM characterization.
Furthermore, PSRs J1022+1001, J1744$-$1134, and even J1012+5307—despite its relatively high ELAT of 38.76$^\circ$—are all affected by the SW signal, which, as discussed in Section~\ref{sec:sw}, biases the DM recovery.
Another, although less-likely, explanation could be found in DM chromaticity \citep{css16}.\\

\subsubsection{Comparison with predicted spectral index}

The spectral indices inferred from both datasets are consistent with the expected Kolmogorov value of $8/3$ (indicated by green vertical lines) for six pulsars: PSRs J0751+1807, J1024$-$0719, J1730$-$2304, J1738+0333, J1857+0943, and J1918$-$0642.
Interestingly, while this was not the case with \texttt{DR2new+}, the spectral index becomes consistent with $8/3$ for PSRs J1012+5307 and J1640+2224, and shows a mild agreement for J1713+0747.
Although the constraints for PSRs J0030+0451, J1022+1001, and J1744$-$1134 remain below the expected value, the distributions for the latter two exhibit an increasing trend. 
While the increase in spectral index from \texttt{DR2new+} to \texttt{DR2low} could be attributed to frequency-dependent ray path averaging, we highlight the significant impact of solar wind effects on these three pulsars. Such effects may bias the estimation of DM variation parameters due to limitations in current solar wind models (cf. Section \ref{sec:sw}), potentially resulting in spectral index constraints below the expected $8/3$ value.
In the context of pulsar timing data, a spectral index lower than the true value typically indicates excess power at higher frequencies in the PSD, which may arise from unmodeled contributions to dispersion variations, such as those induced by the solar wind.\\

\subsubsection{Comparison with NANOGrav 15-yr}

We noticed a clear consistency of the posterior distributions from \texttt{DR2low} with the constraints obtained with NANOGrav 15-yr dataset (black solid lines) for seven of the eleven considered pulsars: PSRs J0030+0451, J1024$-$0719, J1640+2224, J1713+0747, J1738+0333, J1857+0943 and J1918$-$0642, providing an improvement in the robustness of our results. 
The constraints for PSR~J1744$-$1144 are consistent for the spectral index, but \texttt{DR2low} has a lower amplitude. 
We also observed a strong difference for PSR J1012+5307, where NANOGrav 15-yr appears to be more consistent with \texttt{DR2new+}, with higher amplitude and lower spectral index. Such property could be explained by the fact that NANOGrav 15-yr shares the same observing radio frequencies with \texttt{DR2new+} for this pulsar. 
Although, the NANOGrav dataset does not fully overlap in time with \texttt{DR2low} since it contains ToAs from 2004 for this pulsar (starting almost 7 years before \texttt{DR2low}). 
Thus, any effects implying non-stationary DM variations would yield to different spectral properties as well. 
We attributed the large differences for PSRs J1022+1001 and J1730$-$2304 to the short time span of NANOGrav for these two pulsars, with about 5 and 4 years respectively.\\

\subsubsection{Comparison with PPTA DR3}

The posterior distributions from PPTA DR3 (dashed lines) are fully consistent with \texttt{DR2low} for four of the seven overlapping pulsars: PSRs J1024$-$0719, J1713+0747, J1730$-$2304 and J1857+0943, while mildly consistent with PSR~J1744$-$1134, such as for NANOGrav. 
However, the posterior distribution for PSR~J1022+1001 appears to be more consistent with \texttt{DR2new+}, with a lower spectral index than \texttt{DR2low}.
The very poor constraints for PSR~J0030+0451 might be due to the short time span for PPTA data, with about 3 years.

\subsection{Observations for the RN}
\label{ssec:vanillaRN}

\subsubsection{Comparing precision between DR2new+ and DR2low}

We noted a large impact on the RN constraints after adding data from LOFAR and NenuFAR. 
The posterior distributions become significantly narrower for PSRs J1022+1001, J1713+0747, J1730$-$2304, J1744$-$1134 and J1918$-$0642, where even the 3-$\sigma$ contour levels appear now constrained, while it was only the case for PSR~J1713+0747 with \texttt{DR2new+}. 
We also observed an improvement of precision for PSRs J0030+0451, J0751+1807, J1640+2224 and J1857+0943, where a peak appears more prominent in contour levels $\leq 2$-$\sigma$. 
Furthermore, we noted an improvement in the upper limits for PSRs J1024$-$0719 and J1738+0333, for which including LOFAR data helps reducing the covariances between RN and DMv (see the following paragraph for more details). 
We noted a similar precision between both datasets only for PSR~J1012+5307.\\

\subsubsection{Consistency between DR2new+ and DR2low}

The RN distributions from \texttt{DR2low} are fully consistent with \texttt{DR2new+} for PSRs J0030+0451 and J1713+0747, where \texttt{DR2low} results in a more prominent peak.
However, we obtained a constrained RN for PSRs J0751+1807, J1640+2224, J1730$-$2304, J1744$-$1134, J1857+0943 and J1918$-$0642, while it appears fully unconstrained for \texttt{DR2new+}. 
This improved constraint on the RN component is a new result, but it remains highly consistent with \texttt{DR2new+} for PSRs J0751+1807, J1640+2224, J1730$-$2304, and J1857+0943, with tensions of only $0.35\sigma$, $0.06\sigma$, $0.04\sigma$, and $0.18\sigma$, respectively.
It is in mild tension for PSR J1918$-$0642 ($1.02\sigma$) and PSR J1744$-$1134 ($1.32~\sigma$), where \texttt{DR2low} yields a RN highly constrained at low spectral index ($\gamma_{\rm{RN}} \sim 1.1$) and high amplitudes ($\rm{log}_{10} A_{\rm{RN}} \sim -13.3$).
Conversely, while RN was slightly constrained for PSRs J1024$-$0719 and J1738+0333 with \texttt{DR2new+}, it appears significantly reduced with \texttt{DR2low}. 
This behavior could be observed in Figure~\ref{fig:cornerRNDMcorr} (top and bottom plots for PSRs J1024$-$0719 and J1738+0333 respectively), where we observed strong "L-shaped" correlations between RN and DMv, as described in \cite{ffs25} in the case of poor radio frequency coverage. Adding lower frequencies and thus improving the precision on DM variations has removed these correlations, avoiding a false positive for the RN in these cases.
For PSR J1012+5307, the distribution is slightly shifted compared to \texttt{DR2new+}, mainly toward a higher spectral index ($\sim$ from $1$ to $2$), with a mild tension of about $1.46\sigma$.
Interestingly, we observed for the first time a bimodal posterior distribution for the RN of PSR J1022+1001, which lies inside the poorly constrained distribution for \texttt{DR2new+} (tension at only $0.48\sigma$). 
We observed a peak constrained with a low spectral index ($\gamma_{\rm{RN}} \sim 0$), while the second one favors a steeper power law ($\gamma_{\rm{RN}} \sim 5$). 
This feature could reveal a disentanglement between short-term signals with low spectral index, that could be related with pulse shape variations \citep{lkl15,pbc20}, and longer term processes that could be from different origin (e.g., nHz gravitational waves or spin noise). 
It also means that a careful treatment may need to be applied to properly model the RN of this pulsar, as it was already pointed out in \cite{epta23_noise}.
\\

\subsubsection{Comparison with a predicted value}

As described in Section \ref{sec:noise}, the achromatic red noise might have multiple origins (e.g., pulsar spin noise, nanoHertz gravitational waves), which implies some caveats about comparing it against predicted values from single pulsar noise analyses. Nevertheless, it is in practice interesting to apply such comparisons with predicted value for the GWB \citep[see the recent summary by ][]{Verbiest2024}, in particular regarding the slight tensions from the most recent results published by PTAs. 
In this case, we compared the constraints on the spectral index of the RN against the predicted value of $13/3$ from the simplistic case of a GWB caused by a population of GW-driven and circular SMBHBs (green vertical line). 
In the case of \texttt{DR2low}, the spectral index value of $13/3$ is within the $1\sigma$ contour levels for PSRs J1024$-$0719, J1640+2224, J1738+0333, J1857+0943 and J1918$-$0642, and withing the $2\sigma$ contour levels for PSRs J0030+0451 J0751+1807, J1022+1001 (second peak), J1713+0747 and J1730$-$2304. It is however strongly inconsistent with the lower spectral indices of PSRs J1012+5307 and J1744$-$1134 (with 1D marginalized medians resp. equal to $1.75$ and $1.10$), for which the major contribution to the RN might be from another origin (e.g., pulsar-related effects).\\

\subsubsection{Comparison with other PTAs}

As mentioned in Subsection \ref{ssec:vanillaRN}, the short time spans for PSRs J1730$-$2304 and J1022+1001 with NANOGrav 15-yr (about 4 and 5 years respectively), and for PSR~J0030+0451 ($\sim3$ years) with PPTA DR3 are the cause of their large uncertainties. 
We found high consistency in the RN properties between \texttt{DR2low} and other PTAs for PSRs J0030+0451, J1012+5307, J1022+1001 (where only the peak at low spectral index shows up for PPTA), J1024$-$0719, J1640+2224, J1730$-$2304, J1738+0333, J1857+0943 and J1918$-$0642, with tensions lower than $0.6\sigma$. 
We observed mild tensions for PSR J1713+0747 ($1.85\sigma$ and $1.50\sigma$ for NANOGrav and PPTA, respectively), where the constraints of \texttt{DR2low} imply larger amplitudes than the other PTAs. 
We emphasize that the shorter time spans for \texttt{DR2low} (half of the other's datasets) could be the cause of such inconsistencies. 
Given the differences in the timing and noise models employed, we recommend a thorough investigation of the combined datasets to evaluate their consistency for these pulsars in future studies.
For PSR~J1744$-$1134, we observed consistent results with PPTA ($\sim 0.43\sigma$) but a moderate tension with NANOGrav ($\sim 2.18\sigma$) which favors a higher spectral index and a lower amplitude as obtained in \texttt{DR2new+}. 
Inconsistencies between PTAs were already found in \cite{3P+comparison2024} for this pulsar, and hence a combination of the datasets in an IPTA context might help  understand these features better.

\begin{figure}[!htb]
    \centering
    \begin{minipage}{\linewidth} 
        \centering
        \includegraphics[width=\linewidth,height=7.5cm]{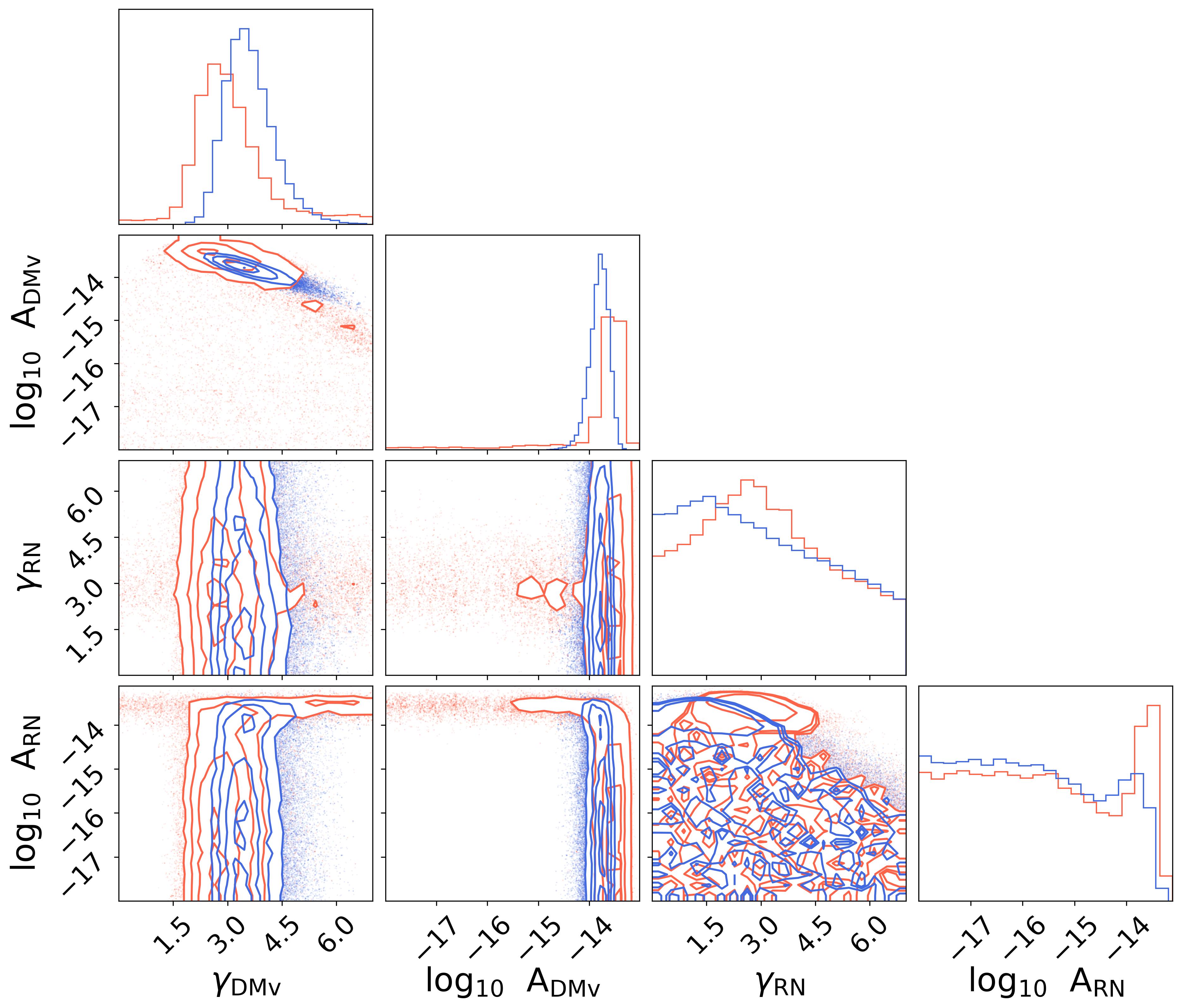}
    \end{minipage}
    \vspace{0.2cm} 
    \begin{minipage}{\linewidth} 
        \centering
        \includegraphics[width=\linewidth,height=7.5cm]{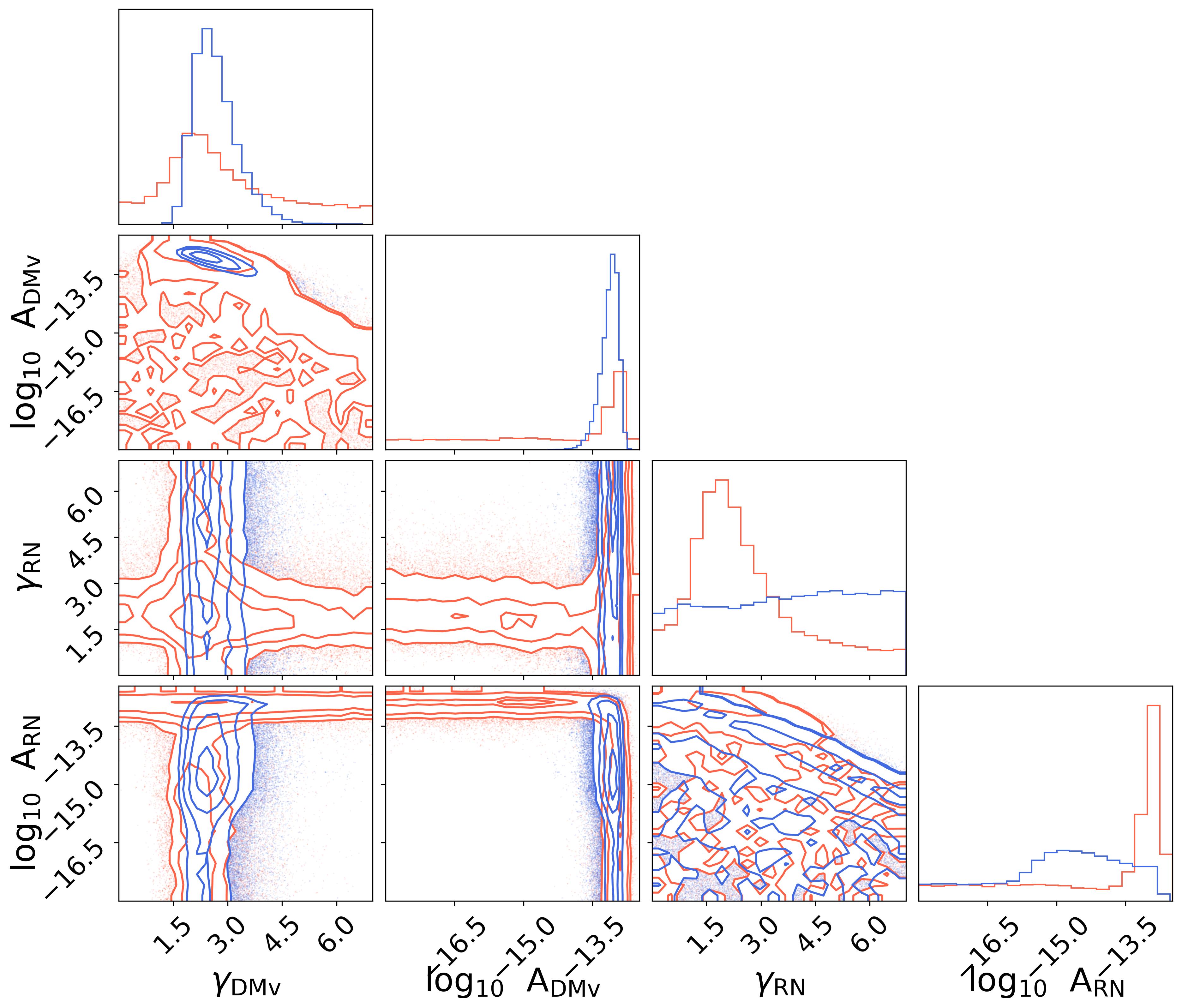}
    \end{minipage}
    \caption{Posterior distributions of spectral indices ($\gamma$) and amplitudes ($A$) of DMv and RN for PSRs J1024$-$0719 (top) and J1738+0333 (bottom) obtained with \texttt{DR2low} (blue) and \texttt{DR2new+} (red), using the standard noise models.}
    \label{fig:cornerRNDMcorr}
\end{figure}

%-----------------------------------------------------------------
\section{Noise analysis - Results from model selection}\label{sec:advanced}
We performed an analysis aimed at finding the noise model that is most favored by the \texttt{DR2low} data, following a similar approach as \cite{epta23_noise}. 
The posterior distributions of \texttt{ECORR} parameters for every pulsars/systems are shown in Figure~\ref{fig:ecorr}, where only five pulsars have unconstrained distributions (PSRs~J0571+1807, J1024$-$0719, J1738+0333, J1857+0943 and J1918$-$0642).
The SW parameters of PSRs J0030+0451, J0751+1807, J1022+1001 and J1744$-$1134 are presented in Figure~\ref{fig:sw parameters}.
In this section, we first intend to provide a comparison with the results that were obtained with \texttt{DR2new+}, then interpret the results for \texttt{DR2low}.

\begin{table*}[ht!]
\centering
\caption{Comparison of favored noise models between \texttt{DR2low} and \texttt{DR2new+}.}
\label{tab:favmod}
% \begin{tabular}{ccccccc}
\begin{tblr}{
  column{1-Z} = {c}, 
  row{1-Z} = {m},  
  cell{1}{2} = {c=3}{c}, 
  cell{1}{5} = {c=3}{c},
  vline{1,2,5,8} = {1-14}{},
  hline{1,2,14} = {-}{},
  rowsep = 1pt
}
\vspace{-.05cm} Pulsar & \vspace{-.05cm} \texttt{DR2low} &        &       & \vspace{-.05cm} \texttt{DR2new+} &        &    \\
J0030+0451             & RN13                            & DMv100 & CN$_4$77  & RN10                             & X      & X  \\
J0751+1807             & X                               & DMv92  & CN$_4$38  & X                                & DMv50  & X  \\
J1012+5307             & RN87                            & DMv20  & CN$_4$100 & RN100                            & DMv47  & X  \\
J1022+1001             & RN98                            & DMv99  & CN$_4$99  & RN30                             & DMv100 & X \\
J1024$-$0719           & X                               & DMv18  & X     & X                                & DMv15  & X \\
J1640+2224             & RN54                            & DMv40  & CN$_4$40  & X                                & DMv47  & X \\
J1713+0747             & RN73                            & DMv74  & CN$_4$12  & RN73                             & DMv72  & X \\
J1730$-$2304           & RN18                            & DMv52  & X     & X                                & DMv11  & X \\
J1738+0333             & RN29                            & DMv28  & X     & X                                & DMv24  & X \\
J1744$-$1134           & RN22                            & DMv100  & CN$_4$31  & RN19                            & DMv98  & X \\
J1857+0943             & RN87                            & DMv99  & X     & X                                & DMv92  & X \\
J1918$-$0642             & RN36                            & DMv24  & CN$_4$42  & X                                & DMv75  & X \\
% \end{tabular}
\end{tblr}
\end{table*}

\subsection{Comparing the favored models of \texttt{DR2new+} and \texttt{DR2low}}

Table \ref{tab:favmod} presents the final noise models obtained with \texttt{DR2low} (left) from this work, and \texttt{DR2new+} (right) from the work in \cite{epta23_noise}. 
It explicitly describes the noise components (RN, DMv and CN$_4$) modeled as GPs with a power-law power spectral density, along with the corresponding favored number of frequency bins.
It is important to note that model selection was not applied to PSR~J1022+1001 for the EPTA DR2 datasets (including \texttt{DR2new+}), as it was found that a single power-law model was insufficient to adequately describe the data across both low and high frequencies. 
Consequently, we excluded this pulsar from the comparison with \texttt{DR2new+}, as the RN30+DMv100 model was retained for this dataset.

We observed a significant change in the final noise models between the two datasets, where the favored models are the same only for PSR J1024$-$0719, with just DMv. 
Firstly, we observed that the RN is favored for 10 of the 12 pulsars with \texttt{DR2low}, while it was only the case for five of them with \texttt{DR2new+}.
This behavior supports the notion that incorporating low-frequency data can enhance PTA sensitivity. 
As demonstrated in \citet{epta23_gwb}, the inclusion of InPTA DR1—featuring observations in the $300$–$500$~MHz range—alongside EPTA DR2 led to a modest increase in the evidence for a GWB signal.
In the case of LOFAR and NenuFAR, we anticipated a stronger improvement since the frequencies are lower ($<200$ MHz), and also because it contains a longer time span and higher cadence, allowing us to better sample signals in the power spectral domain. 
Additionally, we observed the presence of higher order chromatic noise for eight pulsars, while none were observed with \texttt{DR2new+}. 
Lastly, while the DM variations are favored for all pulsars apart from PSR~J0030+0451 with \texttt{DR2new+}, they are now present for all of them with \texttt{DR2low}.
This property is  not surprising, however,  thanks to the high sensitivity of low-frequency data toward DM variations.

For the selected number of frequency bins,  we observed comparable results for the RN component, with relatively low values ($\sim$10–22) for PSRs J0030+0451 and J1744$-$1134, and significantly higher values ($\sim$87–100) for PSRs J1012+5307 and J1713+0747—indicating the presence of constrained power at higher Fourier frequencies.
The outcomes are also similar for the DM variations with PSRs J1024$-$0719, J1640+2224, J1713+0747, J1738+0333, J1744$-$1134 and J1857+0943, while significant differences appear for PSRs J0751+1807, J1012+5307, J1730$-$2304 and J1918+0642.

\subsection{Discussing the results from the noise model selection}
Table~\ref{tab:favparvals} provides the estimated amplitude (set to a reference frequency of 1$\rm{yr}^{-1}$, and $1.4$ GHz for DMv and CN$_4$) and slope (medians with $95\%$ confidence intervals) for the selected components.
The last column displays the Bayes factors in favor of the selected model, against a model composed of \texttt{RN+DMv}, with favored number of frequency bins specifically obtained for this model. 

As described in the previous subsection, the selected models include all the noise components (RN, DMv and CN$_4$) for seven pulsars: PSRs J0030+0451, J1012+5307, J1022+1001, J1640+2224, J1713+5307, J1744$-$1134 and J1918$-$0642.
For these, we observed very high Bayes factors ($>100$) in favor of including CN$_4$ over \texttt{RN+DMv}, apart for PSRs J1713+0747 and J1918$-$0642, with only $\mathcal{B}^{\rm{RN+DMv+CN_4}}_{\rm{RN+DMv}} = 16.0$ and $1.7$ respectively. 
The additional chromatic noise is also included in the favored model for PSR~J0751+1807 (DMv + CN$_4$), where the apparent RN in Figure \ref{fig:corner_vanilla} becomes highly unconstrained once CN$_4$ is included. 
We still noticed low Bayes factors $\mathcal{B}^{\rm{DMv+CN_4}}_{\rm{RN+DMv}} = 9.4$ and $\mathcal{B}^{\rm{DMv+CN_4}}_{\rm{RN+DMv+CN_4}} = 1.2$, indicating only mild support for the final model.
It is similar for PSR J1024$-$0719, where the favored model \texttt{DMv} is poorly favored against \texttt{RN+DMv}: $\mathcal{B}^{\rm{DMv}}_{\rm{RN+DMv+CN_4}} = 1.4$. 
This shows a lack of support for the presence of RN in the data, also being consistent for this pulsar with a similar analysis applied on the PPTA DR2 \citep{grs20}.
For the three remaining pulsars, PSRs J1730$-$2304, J1738+0333 and J1857+0943, the selected models remain composed of \texttt{RN+DMv}.\\

\begin{table*}[ht!]
% \caption{Estimated values (medians with $95\%$ confidence intervals) of the amplitude and spectral index of RN, DMv, and CN$_4$ obtained with \texttt{DR2low} by using the favored noise models displayed in Table \ref{tab:favmod}.}
\caption{Amplitude and spectral index of RN, DMv, and CN$_4$ from \texttt{DR2low}.}
\centering
\label{tab:favparvals}
\begin{tblr}{
  column{1-Z} = {c}, 
  cell{1}{1} = {r=2}{c}, 
  cell{1}{8} = {r=2}{c}, 
  cell{1}{2} = {c=2}{c}, 
  cell{1}{4} = {c=2}{c},
  cell{1}{6} = {c=2}{c},
  vline{1,2,4,6,8,9} = {1-14}{},
  hline{1,3,15} = {-}{},
  hline{2} = {2-7}{}, 
  rowsep = 1pt,
}
Pulsar      & Red noise                &            &  DM variations                &               & Chromatic noise    &          & $\mathcal{B}^{\rm{Fav}}_{\rm{RN+DMv}}$ \\
             & $\rm{log}_{10} \ \rm{A}$ &   $\gamma$ & $\rm{log}_{10} \ \rm{A}$      & $\gamma$      & $\rm{log}_{10} \ \rm{A}$ & $\gamma$ & \\
J0030+0451 & ${-14.79}^{+1.24}_{-2.02}$ & ${4.61}^{+2.26}_{-3.56}$ & ${-13.49}^{+0.10}_{-0.09}$ & ${0.96}^{+0.30}_{-0.27}$ & ${-17.15}^{+0.48}_{-0.77}$ & ${3.12}^{+1.78}_{-1.16}$ & $\geq10^3$ \\
J0751+1807 & X & X & ${-13.37}^{+0.24}_{-0.64}$ & ${2.69}^{+1.72}_{-0.77}$ & ${-15.15}^{+0.17}_{-0.47}$ & ${0.74}^{+1.57}_{-0.68}$ & 9.4 \\
J1012+5307 & ${-12.92}^{+0.14}_{-0.13}$ & ${1.41}^{+0.50}_{-0.48}$ & ${-13.73}^{+0.14}_{-0.12}$ & ${1.72}^{+0.72}_{-0.65}$ & ${-16.30}^{+0.16}_{-0.22}$ & ${1.21}^{+0.53}_{-0.72}$ & $\geq10^3$ \\
J1022+1001 & ${-13.04}^{+0.13}_{-0.12}$ & ${0.15}^{+0.39}_{-0.15}$ & ${-13.27}^{+0.10}_{-0.09}$ & ${1.59}^{+0.34}_{-0.32}$ & ${-16.52}^{+0.20}_{-0.54}$ & ${1.30}^{+1.47}_{-0.67}$ & $\geq10^3$ \\
J1024$-$0719 & X & X & ${-13.74}^{+0.33}_{-0.41}$ & ${3.38}^{+1.51}_{-1.10}$ & X & X & 1.4 \\
J1640+2224 & ${-13.26}^{+0.18}_{-2.27}$ & ${0.28}^{+1.49}_{-0.27}$ & ${-14.26}^{+0.55}_{-1.77}$ & ${3.19}^{+3.71}_{-1.83}$ & ${-15.93}^{+0.19}_{-1.78}$ & ${1.79}^{+4.13}_{-0.86}$ & $\geq10^3$ \\
J1713+0747 & ${-13.87}^{+0.16}_{-0.43}$ & ${2.07}^{+1.83}_{-0.63}$ & ${-13.80}^{+0.11}_{-0.12}$ & ${1.72}^{+0.62}_{-0.50}$ & ${-15.88}^{+0.39}_{-1.23}$ & ${2.30}^{+2.53}_{-1.89}$ & 16.0 \\
J1730$-$2304 & ${-13.83}^{+0.63}_{-1.63}$ & ${3.20}^{+3.46}_{-2.43}$ & ${-13.64}^{+0.30}_{-0.78}$ & ${2.19}^{+2.77}_{-1.21}$ & X & X & X \\
J1738+0333 & ${-15.10}^{+2.21}_{-4.60}$ & ${3.72}^{+3.12}_{-3.52}$ & ${-13.05}^{+0.28}_{-0.51}$ & ${2.45}^{+1.52}_{-0.89}$ & X & X & X \\
J1744$-$1134 & ${-13.34}^{+0.22}_{-0.30}$ & ${1.05}^{+1.20}_{-0.86}$ & ${-13.68}^{+0.10}_{-0.10}$ & ${1.69}^{+0.38}_{-0.32}$ & ${-15.90}^{+0.17}_{-0.25}$ & ${0.53}^{+0.92}_{-0.49}$ & 124.0 \\
J1857+0943 & ${-15.62}^{+1.93}_{-4.15}$ & ${3.72}^{+3.09}_{-3.48}$ & ${-13.31}^{+0.14}_{-0.26}$ & ${1.88}^{+1.11}_{-0.53}$ & X & X & X \\
J1918$-$0642 & ${-15.26}^{+1.90}_{-4.49}$ & ${3.45}^{+3.30}_{-3.21}$ & ${-14.48}^{+1.21}_{-1.15}$ & ${5.03}^{+1.86}_{-3.15}$ & ${-16.33}^{+1.28}_{-3.29}$ & ${3.34}^{+3.43}_{-2.96}$ & 1.7 \\
\end{tblr}
\tablefoot{Median values with $95\%$ confidence intervals are shown. 
The estimates are obtained using the favored noise models listed in Table~\ref{tab:favmod}.}
\end{table*}

\subsubsection{Comparison with standard noise models}

Here we compare our results with those obtained using the standard noise models -- namely \texttt{TM}, \texttt{WN}, \texttt{RN30}, and \texttt{DM100}, as defined in Section~\ref{sec:basic}.
The estimation of \texttt{RN} (if present) and \texttt{DMv} parameters are highly consistent with the ones obtained with the standard noise models (cf. Figure \ref{fig:corner_vanilla}) for PSRs J0030+0451, J0751+1807, J1024$-$0719, J1713+0747, J1730$-$2304, J1738+0333, J1744$-$1134 and J1857+0943. 
Mild differences are observed for PSR~J1012+5307, with DMv being slightly shifted toward a lower spectral index, and PSR~J1918$-$0642, where the RN and DMv are respectively less and more constrained. 
For the case of PSR~J1022+1001, while DMv remains fully consistent, the RN becomes highly constrained to the peak at the low spectral index (cf. Figure \ref{fig:corner_vanilla}), due to the inclusion of many frequency bins (\texttt{RN98}). 
However, we emphasize that such noise model is expected to be insufficient to properly include the power of the RN at low PSD frequencies. 
A dedicated study for this pulsar is still required to ensure a proper characterization of the noise. 
Lastly,  we observed significant changes on both RN and DMv parameter posterior distributions for PSR~J1640+2224, where the RN gets highly constrained at a low spectral index and high amplitude, while the DMv displays a bimodal shape, with a main peak consistent with the standard noise model and a secondary peak favoring high spectral indices ($\gamma>7$) and low amplitudes ($\rm{log}_{10} \rm A < -15$).\\

\subsubsection{Examining the value of the chromatic index}

\begin{figure}[ht!]
    \centering
    \includegraphics[width=\linewidth]{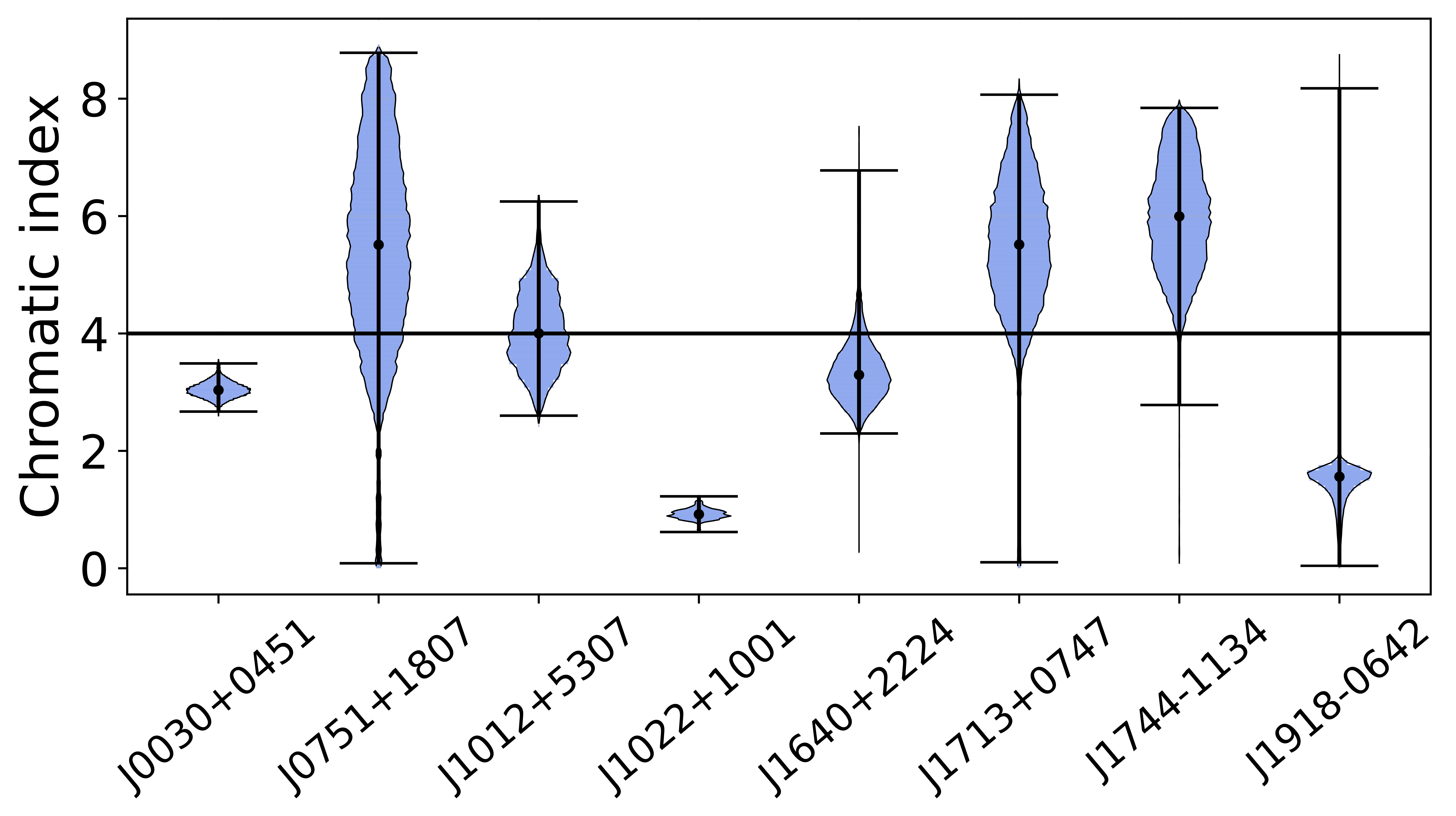}
    \caption{Posterior distribution of the chromatic index (blue violins) and their $99.7\%$ credible intervals (errorbars) obtained for the pulsars that favor the presence of CN$_4$ in their noise model. 
    Here we use the favored noise model and replace the CN$_4$ noise component by a Gaussian process power-law PSD signal with the same number of frequency bins, but a free chromatic index, with a prior defined as $\mathcal{U}(0,10)$. The black horizontal line displays the fixed value for the case of CN$_4$.}
    \label{fig:ChromaticNoise}
\end{figure}

To validate the choice of the chromatic index of CN$_4$, we performed an additional parameter estimation using the same model as in Table~\ref{tab:favmod}, but instead of fixing its chromatic index, we treat it as a free parameter $\chi$.
Figure~\ref{fig:ChromaticNoise} displays the posterior distributions of the chromatic indices obtained for all of the pulsars that include CN$_4$ in the noise model and the error bars display the $99.7\%$ credible interval of the distribution. 
We observed a strong consistency with the predicted value of $4$ from the thin screen approximation (shown as a black horizontal line in the figure) for PSR~J1012+5307, with $\chi = 4.00_{-0.6}^{+0.8}$ (median with $68\%$ credible intervals). 
The results are also consistent with this predicted value for PSRs J0751+1807, J1713+0747 and J1744$-$1134, that all display constraints favoring higher values, and also for PSR~J1640+2224, where the posterior distribution is favoring slightly lower values. 
Instead, the chromatic index of PSR~J1918$-$0642 appears constrained at lower values, with a main peak centered at $\chi = 1.6_{-0.4}^{+0.4}$. 
However, we also observed a secondary peak centered at 4, leading to a $99.7\%$ credible interval consistent with the predicted value. 
Thus, we suggest conducting a dedicated analysis of this pulsar, particularly when incorporating low radio frequency data. 
The two remaining pulsars, PSRs J0030+0451 and J1022+1001 provide chromatic indices that are highly inconsistent with thin screen turbulence theory, being respectively $\chi = 3.0_{-0.1}^{+0.1}$ and $0.92_{-0.08}^{+0.09}$. 
We emphasize the presence of strong dispersion variations caused by the solar wind for both pulsars, which would induce a reduction of the measured value in case of imperfect modeling (cf. next section) since DMv has a chromaticity of $2$. 
Furthermore, we could also expect a lower value for PSR~J1022+1001 as we pointed a bimodal distribution of RN parameters that is poorly modeled with a single power law. 
Thus, an imperfect treatment of RN, for which the chromaticity is equal to zero, would lead to reduction of the measured chromatic index.

\section{Impact of solar wind}\label{sec:sw}

In this section, we study in more detail the impact of the SW model on PSRs J0030+0451 and J1022+1001.
Specifically, we aim to investigate the differences in the noise properties when the SW signature is absent by inspecting the free spectra of DM variations and RN, where we fit for the power spectral density amplitude for each bin independently.
For this reason, we analyzed two truncated datasets per pulsar by removing the \texttt{DR2low} data taken closer than 45$^\circ$ and 75$^\circ$ to the Sun, respectively.
The 45$^\circ$ threshold follows \citet{tvs19}, where it is shown that the SW effects are well modeled beyond this angular distance from the Sun (see their Figure~12).
The more conservative 75$^\circ$ cut is introduced because NenuFAR— whose data were absent in \citet{tvs19}—offers an even higher sensitivity to DM effects than LOFAR.
This stricter cut ensures that any residual and unmodeled SW contribution is removed, even in the presence of these highly sensitive datasets.

\begin{figure*}
    \caption{Free spectra for DMv (referred to 1.4GHz) and RN for PSR~J0030+0451 (top) and PSR~J1022+1001 (bottom).
    Each violin represents the posterior distribution of the RMS amplitude in a given Fourier bin. 
    The blue violins correspond to the full \texttt{DR2low} dataset; the green and orange violins show the results obtained after applying solar angle cuts of 45° and 75°, respectively.
    The vertical dashed lines indicate the harmonics of the $1\mathrm{yr^{-1}}$ frequency, which is representative of the SW effect.
    The number of Fourier bins displayed matches that of the preferred noise model listed in Table~\ref{tab:favmod}.
    }
    \centering
    \includegraphics[width=\textwidth,height=6cm]{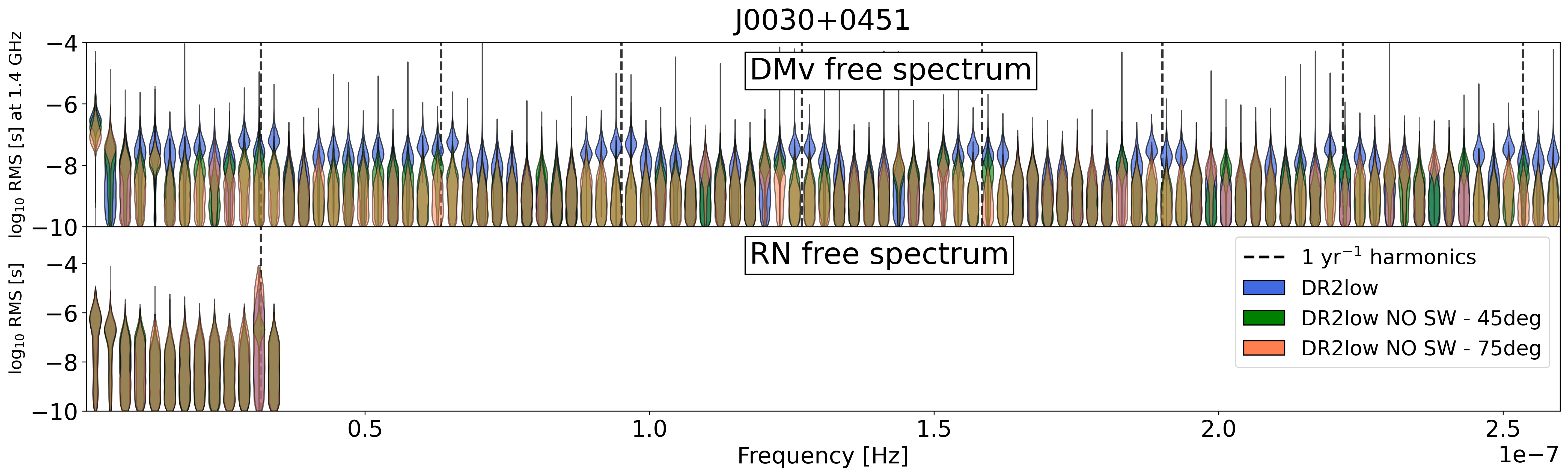}
    \includegraphics[width=\textwidth,height=6cm]{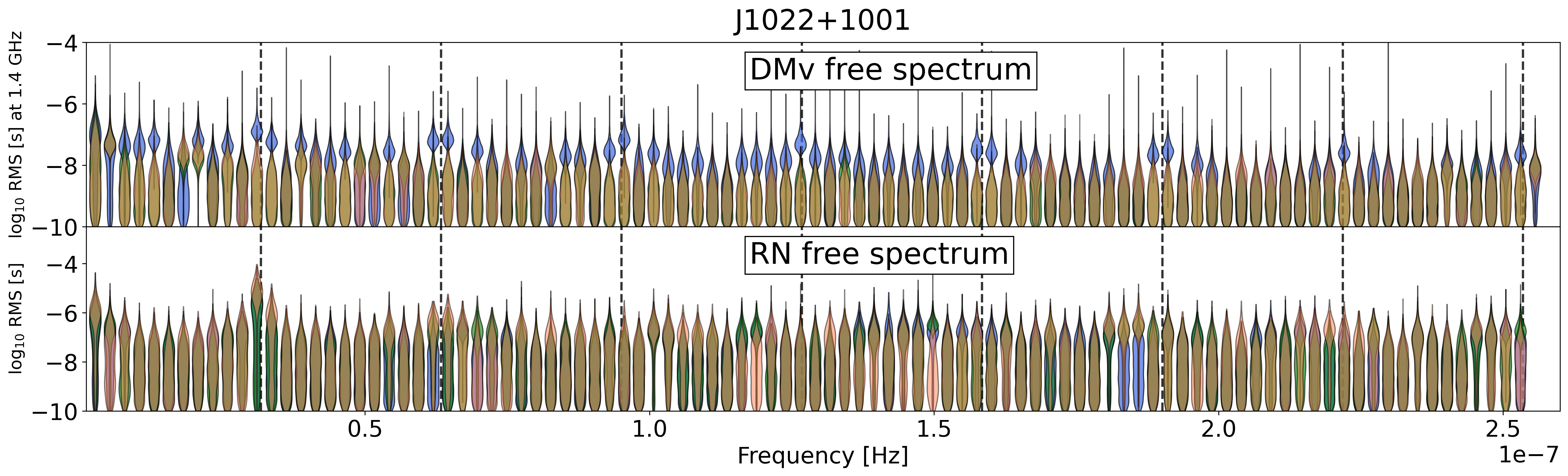}
    \label{fig:dmfreespec}
\end{figure*}

The adopted noise model is the same as in Table~\ref{tab:favmod}; however,  in the previous analysis the SW was mitigated following Table~\ref{tab:SWmodelling}, while in this case it is no longer part of the model.
In Figure~\ref{fig:dmfreespec} we show the resulting free-spectra.
By inspecting the DM spectra of the \texttt{DR2low} we observed that there is a noticeable increase in power at the harmonics of $1\mathrm{yr}^{-1}$ compared to the two cases with solar angle cuts.
This suggests that the adopted SW model does not fully capture the entirety of the signal dynamics leading to a periodic leakage into the DM spectrum.
Hence, the simple power-law model compensates by flattening its spectral index and increasing the number of Fourier components as shown in Section~\ref{sec:basic} and \ref{sec:advanced} for the solar pulsars such as PSRs J0030+0451, J1022+1001 and J1744$-$1134.
This compensatory behavior highlights the limitations of the current SW modeling in accurately characterizing its induced noise.
In future works one could explore some improvement of the current SW model by changing the number of Fourier bins and the power-law assumption of the PSD.
We also noticed that the $45^\circ$- and $75^\circ$-cut datasets yield consistent results across almost all bins in the spectrum. 
In contrast, the RN does not appear to be affected by the SW since no differences in the free-spectra are detected among the three datasets.

\section{Conclusions}\label{sec:conclusion}

In this work we combined high-precision timing data from the EPTA DR2new+ with low-frequency observations from LOFAR and NenuFAR.
Among the 25 MSPs included in the DR2new+, we identified respectively 12 and 2 sources in common with the LOFAR and NenuFAR datasets.
The time span of the pulsars in common almost perfectly overlaps with that of \texttt{DR2new+}.
The addition of low-frequency radio data significantly extends the frequency coverage of the dataset, enhancing our sensitivity to propagation effects such as DM variations and higher order chromatic delays.
This new dataset is referred to as \texttt{DR2low}.
We performed multiple iterations of the noise analysis, following the approach outlined in \citet{epta23_noise}, and we compared the results obtained with \texttt{DR2low} against those from \texttt{DR2new+}.
As a first step, we adopted a \texttt{standard noise model} consisting of the \texttt{TM}, \texttt{WN}, \texttt{RN30}, and \texttt{DMv100} components.
In addition, we applied SW corrections to those pulsars with low ecliptic latitudes and sufficient observing cadence around solar conjunctions (see Table~\ref{tab:SWmodelling}).
The corner plots for the \texttt{standard noise model} are presented in Figure~\ref{fig:corner_vanilla}.

Overall, the results show a significant improvement in constraining the DM variations, as expected with the increased low-frequency coverage of the dataset. 
Moreover, this enhanced sensitivity allows us to more clearly distinguish between DM variations and RN, or at least to reduce the correlation between the two processes. 
This improvement is evident in Figure~\ref{fig:cornerRNDMcorr}, where the typical L-shaped correlation between RN and DMv has been mitigated by the inclusion of low-frequency data.
This result is consistent with the simulations presented by \citet{ffs25}.
Furthermore, we showed an improved consistency with the results from NANOGrav 15y and PPTA DR3.

We then performed a more in-depth analysis to find the favored noise model for \texttt{DR2low}.
We observed a significant change in the final noise model with respect to \texttt{DR2new+} (see Table~\ref{tab:favmod}).
The extended time span and high cadence of the low-frequency data (approximately 10 years) now allow us to better sample signals across the power spectral domain.
This improvement enables us to accurately redistribute power into the correct noise components.
As a result, we found that ten pulsars have significant RN in the final model in contrast to the five in \texttt{DR2new+}.
Moreover, DMv are present in all the 12 pulsars, while for \texttt{DR2new+} this component was missing for PSR~J0030+0451.
We also found an additional noise component with a fixed chromatic index of 4 in the final models of 8 out of 12 pulsars, whereas none were detected in \texttt{DR2new+}. 
To validate the chosen chromatic index, we fit for this parameter ($\chi$) in an additional analysis.
Five out of eight pulsars show consistency with the expected value from a thin screen with same-size inhomogeneities, $\chi = 4$.
A more detailed analysis is required for PSR~J1918$-$0642, which displays a main peak at $\chi < 2$, but also a secondary peak that is consistent with $\chi=4$.
For PSRs J0030+0451 and J1022+1001 the chromatic index is inconsistent with 4, possibly due to the strong DM variations caused by the solar wind.

Finally, we conducted a detailed study on the impact of the SW model for PSR~J0030+0451 and PSR~J1022+1001. 
We applied two thresholds to remove epochs from \texttt{DR2low} with solar angles smaller than 45$^\circ$ and 75$^\circ$, respectively. 
We analyzed the datasets using the same final noise model shown in Table~\ref{tab:favmod}, while excluding the SW model in the truncated datasets.
Figure~\ref{fig:dmfreespec} shows the free spectra of DMv and RN. 
We observed clear excess power in the DMv around the harmonics of the $1\mathrm{yr}^{-1}$ frequency.
This suggests that the current SW model does not fully capture all the features of this process, causing the excess power to be absorbed into the DMv modeling.

The inclusion of low-frequency data has proven to significantly improve the noise modeling in PTAs. 
These data help  disentangle noise components and offer enhanced sensitivity to DM variations and other chromatic signals.
Moreover, the addition of low-frequency observations has been shown to improve the accuracy of astrometric parameters, as demonstrated in \citet{lpk25}. 
This initial integration of LOFAR and NenuFAR data is a crucial step, as both datasets will be included in the upcoming IPTA DR3 \citep{Good2023}, which is set to become the most extensive PTA dataset ever created, with the widest frequency coverage.
Furthermore, LOFAR and NenuFAR will also be part of the next EPTA Data Release 3, which will feature a larger sample of pulsars in common between the datasets.
In this context, the present analysis lays important groundwork for future studies. 
In particular, it will be especially interesting to assess the effect of these very low-frequency data on the detection and characterization of a GWB in upcoming data releases. 
The extended frequency coverage provided by LOFAR and NenuFAR has the potential to improve sensitivity to the GWB and will be an important aspect to explore in future work.

Finally, it is important to highlight the significance of future low-frequency observations, particularly with SKA-low. 
The expected improvements in sensitivity and time coverage provided by SKA-low will be indispensable for advancing pulsar timing array studies and for ensuring the continued success of gravitational wave detection efforts.

\begin{acknowledgements}
FI is supported by the University of Cagliari (IT). FI, CT, AP are supported by the Istituto Nazionale di Astrofisica.
%Aurelien aknowledgement
JPWV acknowledges support from NSF AccelNet award No. 2114721.
SCS acknowledges the support of a College of Science and Engineering University of Galway Postgraduate Scholarship.
GMS acknowledge financial support provided under the European Union’s H2020 ERC Consolidator Grant “Binary Massive Black Hole Astrophysics” (B Massive, Grant Agreement: 818691).
J.A. acknowledges support from the European Commission under project ARGOS-CDS (Grant Agreement number: 101094354)
MB acknowledges the support from the Department of Atomic Energy,
Government of India through Apex-I Project - Advance Research and Education in Mathematical Sciences at IMSc.
This paper is partially based on data obtained using the NenuFAR radio-telescope. The development of NenuFAR has been supported by personnel and funding from: Observatoire Radioastronomique de Nan\c{c}ay, CNRS-INSU, Observatoire de Paris-PSL, Université d’Orléans, Observatoire des Sciences de l’Univers en Région Centre, Région Centre-Val de Loire, DIM-ACAV and DIM-ACAV + of Région Ile-de-France, Agence Nationale de la Recherche. 
We acknowledge the use of the Nan\c{c}ay Data Centre computing facility (CDN – Centre de Données de Nan\c{c}ay). The CDN is hosted by the Observatoire Radioastronomique de Nan\c{c}ay (ORN) in partnership with Observatoire de Paris, Université d'Orléans, OSUC, and the CNRS. The CDN is supported by the Région Centre-Val de Loire, département du Cher. 
The Nan\c{c}ay Radio Observatory (ORN) is operated by Paris Observatory, associated with the French Centre National de la Recherche Scientifique (CNRS) and Universit\'{e}
d'Orl\'{e}ans.
LOFAR, the Low-Frequency Array designed and constructed
by ASTRON, has facilities in several countries, that
are owned by various parties (each with their own funding
sources), and that are collectively operated by the International
LOFAR Telescope (ILT) foundation under a joint scientific policy.
BCJ acknowledges the support from the Raja Ramanna Chair fellowship of the Department of Atomic Energy, Government of India (grant 3/3401 Atomic Energy Research 00 004 Research and Development 27 02 31 1002//2/2023/RRC/R\&D-II/13886) and  the support of the Department of Atomic Energy, Government of India, under Project Identification No. RTI 4002 and  under project No. 12-R\&D-TFR-5.02-0700.
IK acknowledges the support of Coll\`ege de France by means of “PAUSE -Solidarit\'e Ukraine” program and he would like to thank the Paris Observatory and the CNRS (LPC2E) for being great host organizations for the PAUSE program.
This work is supported by 100101 Key Laboratory of Radio Astronomy and Technology (Chinese Academy of Science). 
AC and APa acknowledges financial support from the European Research Council (ERC) starting grant 'GIGA' (grant agreement number: 101116134) and through the NWO-I Veni fellowship. 
DJS acknowledges support by the project ‘‘NRW-Cluster for data intensive
radio astronomy: Big Bang to Big Data (B3D)’’ funded through the programme ‘‘Profilbildung 2020’’, an initiative of the Ministry of Culture
and Science of the State of North Rhine-Westphalia.
KT is partially supported by JSPS KAKENHI Grant Numbers 20H00180, 23K20868, and 21H04467, and Bilateral Joint Research Projects of JSPS.
JW acknowledges support by the BMBF Verbundforschung under the grant 05A23PC2.
JS acknowledges the support from the University of Cape Town Vice Chancellor’s Future Leaders 2030 Awards programme and the South African Research Chairs Initiative of the Department of Science and Technology and the National Research Foundation.
DD acknowledges the Department of Atomic Energy, Government of India’s support through ‘Apex Project-Advance Research and Education in Mathematical Sciences’ at The Institute of Mathematical Sciences. 
PR acknowledges the financial assistance of the South African Radio Astronomy  Observatory (SARAO) towards this research (www.sarao.ac.za).

\end{acknowledgements}

\bibliographystyle{aa}
\bibliography{./bibliography}
\onecolumn
\begin{appendix}

\section{Parameters of the noise models used in this work}\label{Appendix data}

\begin{table}[ht!]
% \caption{List of the noise components used in this work and their corresponding abbreviations used in this document (first column) displayed with their parameter names (second column) and prior definitions.}
\caption{Noise components and corresponding abbreviations.}
\centering
\label{tab:parprior}
\begin{tblr}{
  column{1-Z} = {c}, 
  cell{2}{1} = {r=3}{c}, 
  cell{5}{1} = {r=2}{c}, 
  cell{7}{1} = {r=2}{c}, 
  cell{9}{1} = {r=2}{c}, 
  cell{11}{1} = {r=4}{c},
  cell{15}{1} = {r=3}{c},
  vline{1,2,3,4,5,6,7} = {-}{},
  hline{1,2,5,7,9,11,15,18} = {-}{},
}
Noise component (abrev.)   & Parameters          & Priors                                           \\
White noise (WN)           & \texttt{EFAC}                & $\mathcal{U}(0.01,10)$                           \\
                           & \texttt{EQUAD}               & $\rm{log}_{10} \ \mathcal{U}(10^{-9},10^{-5})$   \\
                           & \texttt{ECORR}               & $\rm{log}_{10} \ \mathcal{U}(10^{-9},10^{-2})$   \\
Achromatic red noise (RN)  & $A_{\rm{RN}}$       & $\rm{log}_{10} \ \mathcal{U}(10^{-20},10^{-11})$ \\
                           & $\gamma_{\rm{RN}}$  & $\mathcal{U}(0,7)$                               \\
DM variations (DMv)        & $A_{\rm{DMv}}$      & $\rm{log}_{10} \ \mathcal{U}(10^{-20},10^{-11})$ \\
                           & $\gamma_{\rm{DMv}}$ & $\mathcal{U}(0,7)$                               \\
Chromatic noise (CN$_4$) & $A_{\rm{CN_4}}$       & $\rm{log}_{10} \ \mathcal{U}(10^{-20},10^{-11})$ \\
                           & $\gamma_{\rm{CN_4}}$  & $\mathcal{U}(0,7)$                               \\
Exponential dip (E)        & $A_E$               & $\rm{log}_{10} \ \mathcal{U}(10^{-10},10^{-2})$  \\
                           & $\tau_E$            & $\rm{log}_{10} \ \mathcal{U}(10^0,10^{2.5})$             \\
                           & $t_{0,E}$           & $\mathcal{U}(57490.0,57530.0)$                   \\
                           & $\chi_E$            & $\mathcal{U}(0,10)$                              \\
Solar wind dispersion (SW) & $\rm{n_{earth}}$    & $\mathcal{U}(0,30)$  \\
                           & $A_{\rm{SW}}$       & $\rm{log}_{10} \ \mathcal{U}(10^{-10}, 10^{1})$        \\
                           & $\gamma_{\rm{SW}}$  & $\mathcal{U}(-2,-1)$          \\
\end{tblr}
\tablefoot{The table lists the noise components used in this work with their abbreviations (first column), parameter names (second column), and prior definitions (third column).
$\mathcal{U}(a,b)$ denotes a uniform prior between $a$ and $b$.}
\end{table}

\begin{table}[ht!]
\centering
\caption{Modeling approach for dispersion caused by solar winds.}
% \caption{Approach for modeling the dispersion caused by solar winds inspired by \cite{Susarla2024}. The pulsar names and ecliptic latitudes (in degrees) are displayed respectively in the first and second columns. The methods are described for \texttt{DR2low} (third major column) and \texttt{DR2new+} (last major column), where the timing analysis include the constant term only (either fixed of fitted with \texttt{Tempo2}), and where the noise analysis includes the constant term, by marginalizing over its error (marg.; see Section \ref{sec:noise}) or fitting for it ($\rm{n_{earth}}$), and eventually adds its time-varying component ($\rm{SWv}$), where the fluctuations of the electron density at one astronomical unit are modeled as a power-law PSD parameterized with its amplitude $A_{\rm{SW}}$ referred to $f=1 \ \rm{yr}^{-1}$ and spectral index $\gamma_{\rm{SW}}$.}
\begin{tblr}{
  column{1-Z} = {c},
  cell{1}{1} = {r=2}{c}, 
  cell{1}{2} = {r=2}{c}, 
  cell{1}{3} = {c=2}{c},
  cell{1}{5} = {c=2}{c},
  vline{1,2,3,5,7} = {1-14}{},
  vline{4,6} = {2-14}{}, 
  hline{1,3,15} = {-}{},
  hline{2} = {3-5}{},
  hline{2} = {5-7}{},
  rowsep = 1pt,
}
Pulsar       & ELAT (deg.)              & SW modeling $-$ \texttt{DR2low} &                                  & SW modeling $-$ \texttt{DR2new+} &      \\
             &                          & Timing - NE\_SW                          & Noise                            & Timing - NE\_SW                            & Noise\\
J1022+1001   & \hspace{-0.3cm} $-0.06$  & $9.51 \pm 0.03$ (fitted)          & \texttt{n\_earth} + \texttt{SWv} & $7.2 \pm 0.5$ (fitted)           & marg. \\
J1730$-$2304 & $0.19$                   & $9.9 \pm 0.4$ (fitted)          & \texttt{n\_earth}                & 7.9 (fixed)                       & marg. \\
J0030+0451   & $1.45$                   & $8.20 \pm 0.03$ (fitted)          & \texttt{n\_earth} + \texttt{SWv} & 7.9 (fixed)                       & marg. \\
J0751+1807   & \hspace{-0.3cm} $-2.81$  & 7.9 (fixed)                      & \texttt{n\_earth} + \texttt{SWv} & 7.9 (fixed)                       & marg. \\
J1744$-$1134 & $11.81$                  & $3.62 \pm 0.08$ (fitted)          & \texttt{n\_earth} + \texttt{SWv} & 7.9 (fixed)                       & marg. \\
J1918$-$0642 & $15.35$                  & 4 (fixed)                        & marg.                             & 7.9 (fixed)                       & marg. \\
J1024$-$0719 & \hspace{-0.3cm} $-16.04$ & 4 (fixed)                        & marg.                             & 7.9 (fixed)                       & marg. \\
J1738+0333   & $26.88$                  & 4 (fixed)                        & marg.                             & 7.9 (fixed)                       & marg. \\
J1713+0747   & $30.70$                  & 4 (fixed)                        & marg.                             & 7.9 (fixed)                       & marg. \\
J1857+0943   & $32.32$                  & 4 (fixed)                        & marg.                             & 7.9 (fixed)                       & marg. \\
J1012+5307   & $38.76$                  & 4 (fixed)                        & marg.                             & 7.9 (fixed)                       & marg. \\
J1640+2224   & $44.06$                  & 4 (fixed)                        & marg.                             & 7.9 (fixed)                       & marg. \\
\end{tblr}
\tablefoot{The modeling strategy follows \citet{Susarla2024}. 
The pulsar names and ecliptic latitudes (in degrees) are listed in the first and second columns. 
For both \texttt{DR2low} (third major column) and \texttt{DR2new+} (last major column), 
the table shows the adopted methods for timing and noise analyses. 
In the timing analysis, the constant term is either fixed or fitted with \textsc{TEMPO2}. 
In the noise analysis, the constant term is included by either marginalizing over its error (marg.; see Section~\ref{sec:noise}) or fitting for it ($\rm{n_{earth}}$). 
The time-varying component ($\rm{SWv}$) represents fluctuations in the electron density at 1~AU, 
modeled as a power-law PSD parameterized by its amplitude $A_{\rm{SW}}$, referred to $f=1~\mathrm{yr}^{-1}$, and spectral index $\gamma_{\rm{SW}}$.}
\label{tab:SWmodelling}
\end{table}

\newpage
\section{Estimation of the \texttt{ECORR} parameters}\label{Appendix ecorr}

\begin{figure*}[ht!]
    \centering
    \caption{Posterior distribution of the \texttt{ECORR} parameters (logarithmic scale) and their $99.7\%$ credible intervals (errorbars) obtained for the LOFAR systems for each pulsar.}
    \includegraphics[width=\linewidth-.5cm, height=22cm]{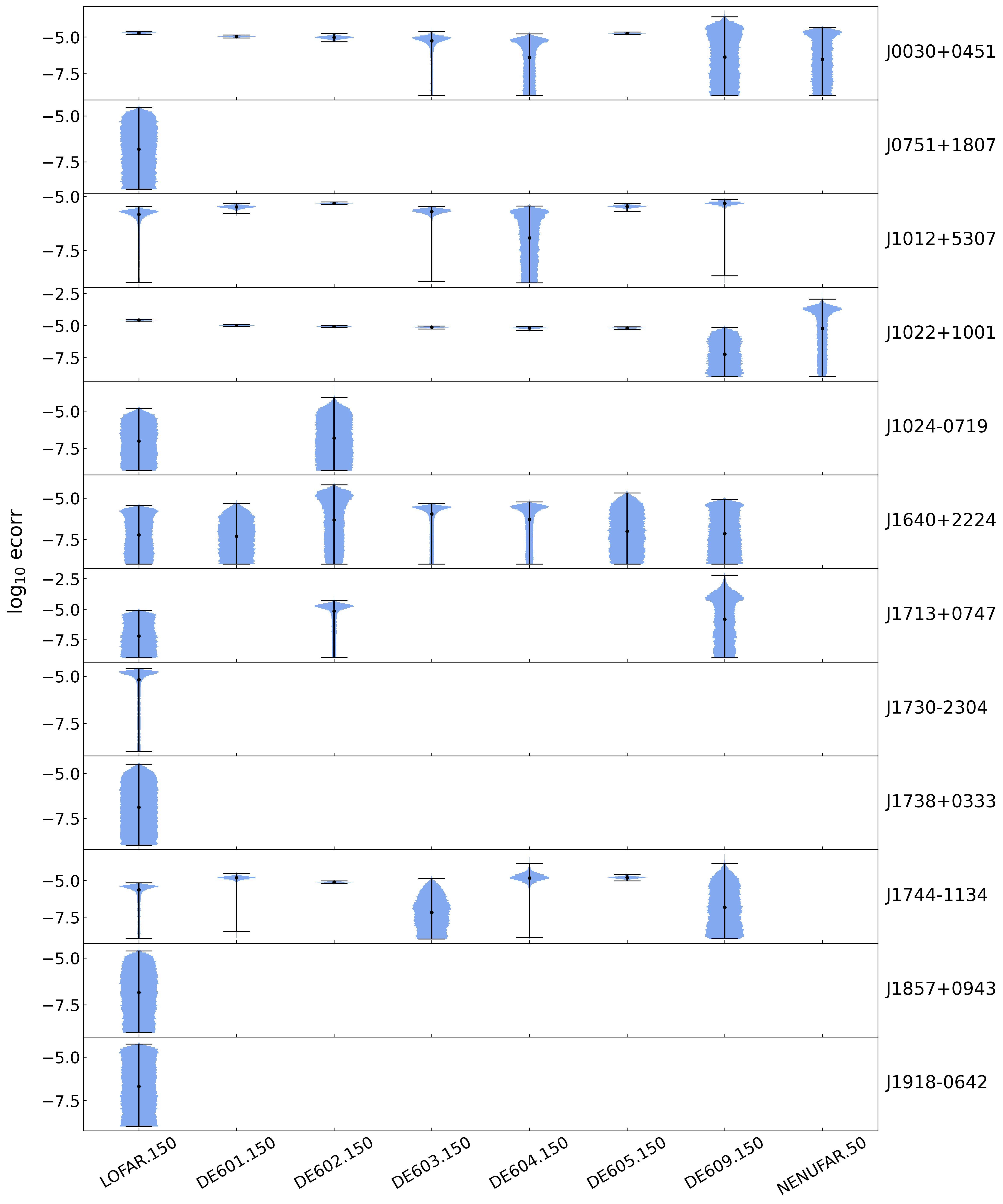}
    \label{fig:ecorr}
\end{figure*}

\newpage
\section{Solar wind parameters}\label{appendix: SW parameters}

\begin{figure}[h!]
\caption{Posterior distribution of the SW parameters for PSRs J0030+0451, J0751+1807, J1022+1001, and J1744$-$1134.}
\vspace{0.3cm}
    \centering
    \begin{minipage}[t]{0.48\linewidth}
        \centering
        \includegraphics[width=\linewidth]{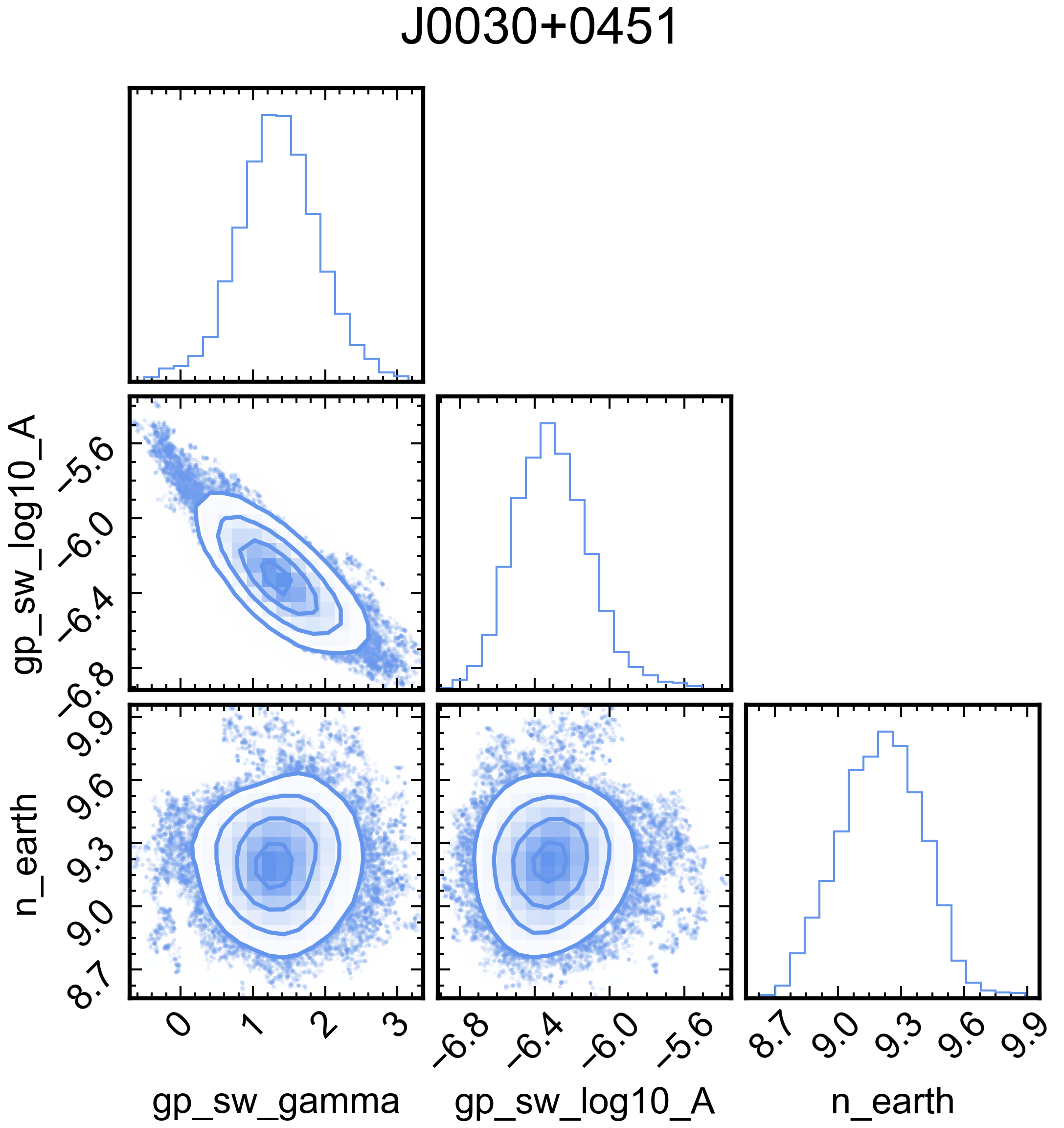}
        \label{fig:sw parameters}
    \end{minipage}
    \hfill
    \begin{minipage}[t]{0.48\linewidth}
        \centering
        \includegraphics[width=\linewidth]{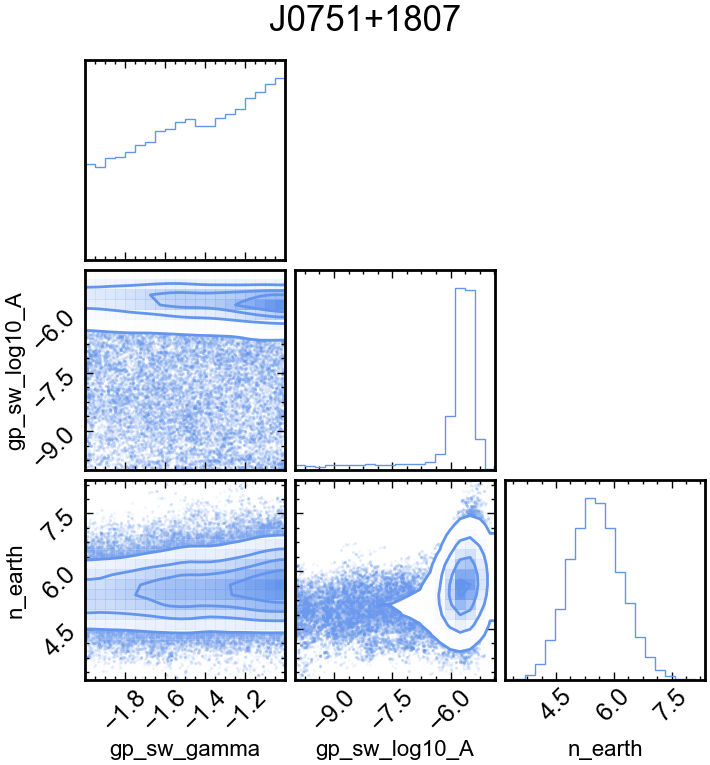}
    \end{minipage}
    
    \vspace{0.5cm}
    
    \begin{minipage}[t]{0.48\linewidth}
        \centering
        \includegraphics[width=\linewidth]{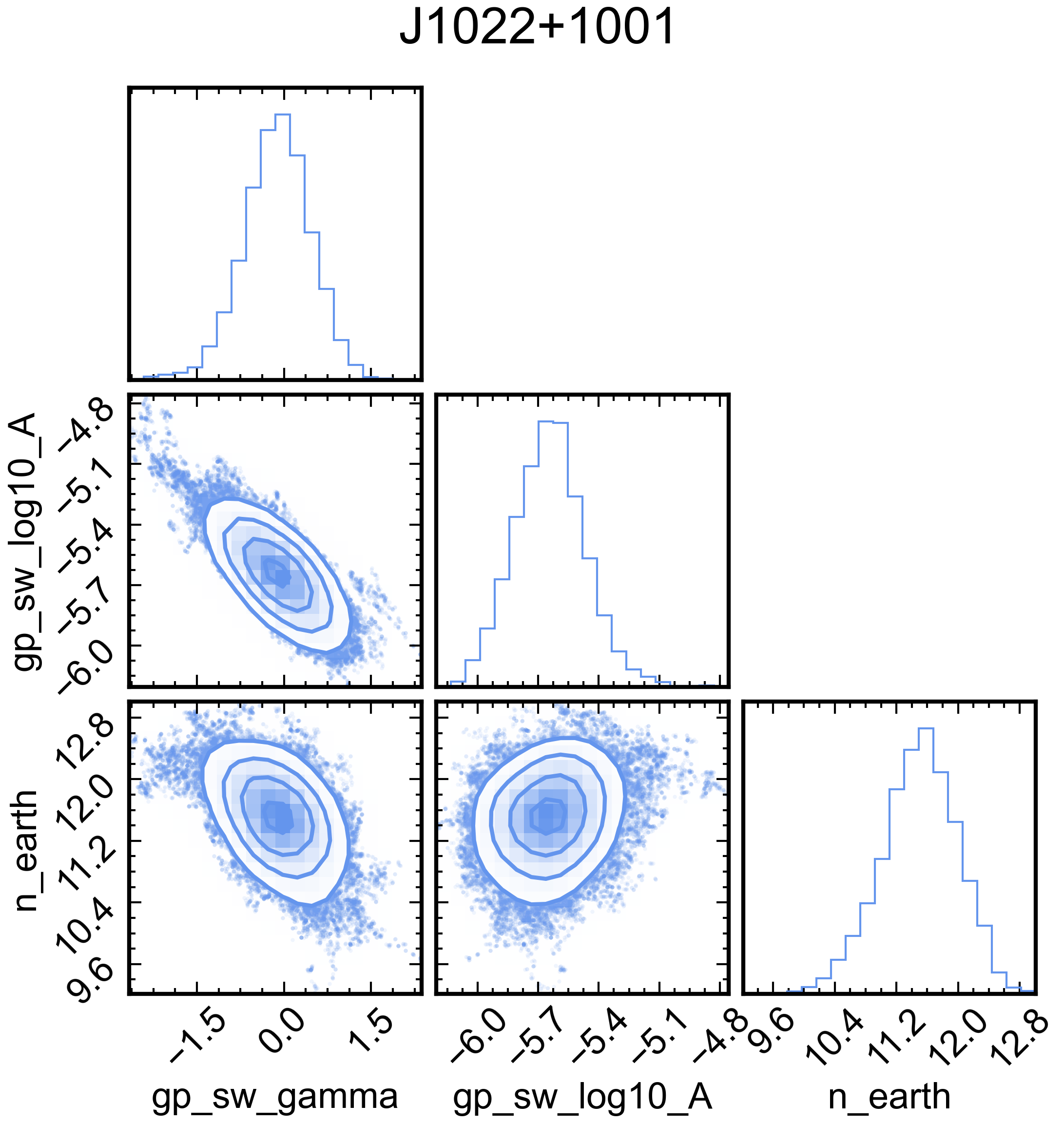}
    \end{minipage}
    \hfill
    \begin{minipage}[t]{0.48\linewidth}
        \centering
        \includegraphics[width=\linewidth]{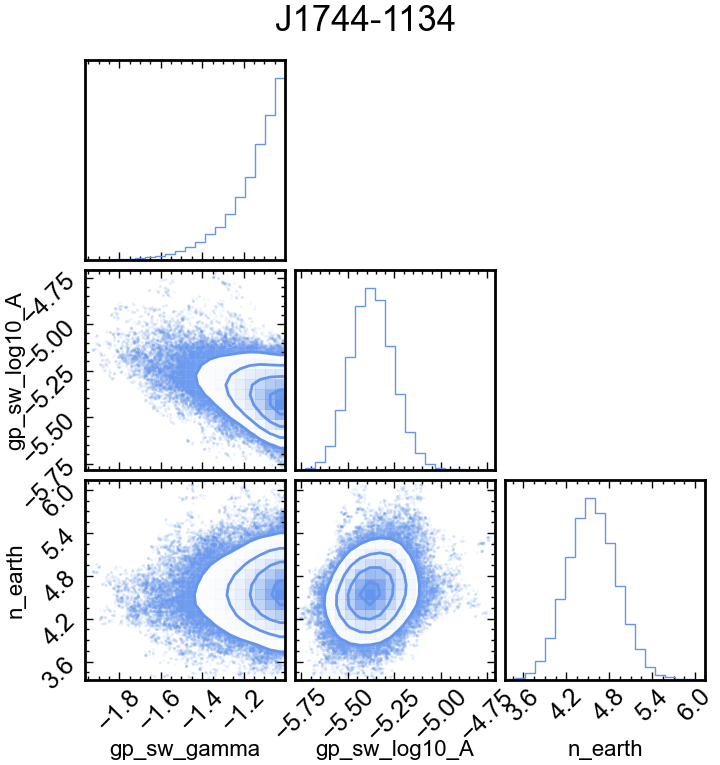}
    \end{minipage}
\end{figure}

\newpage
\section{Timing residuals and observing radio frequencies of \texttt{DR2low}}\label{Appendix data}

\begin{figure*}[ht!]
    \centering
    \caption{Time in year and MJD (resp. for the top and bottom x-axes) vs. observing frequency (in MHz) shown in logarithmic scale for the 12 pulsars considered in this work displayed in green, orange, and blue respectively  for the \texttt{DR2new+}, LOFAR, and NenuFAR data. The gray vertical areas emphasize the periods when the pulsar-sun angles are equal to or lower than 20 degrees.}
    \includegraphics[width=\linewidth-.5cm, height=22cm]{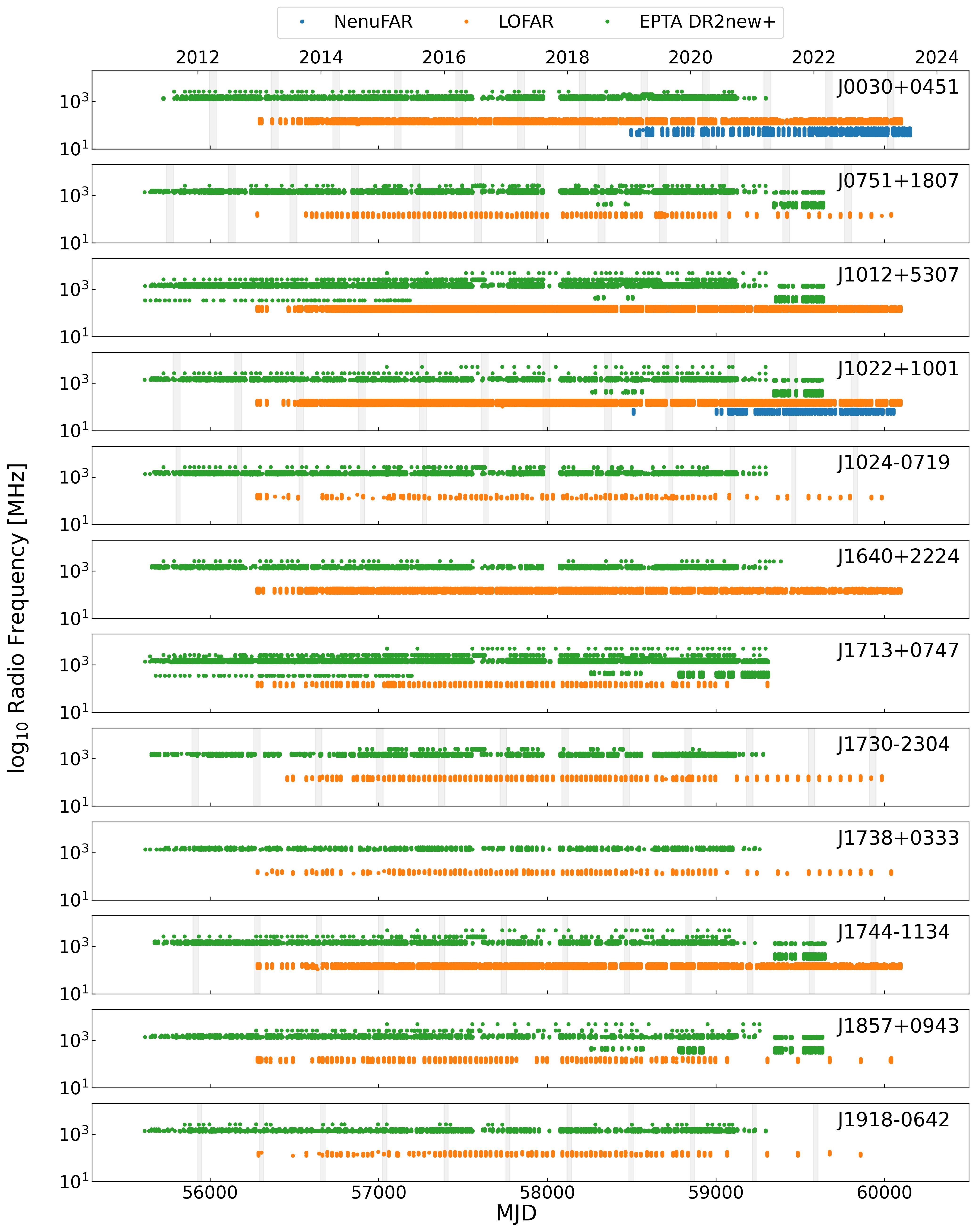}
    \label{fig:t_vs_freqs}
\end{figure*}

\begin{figure*}[ht!]
    \centering
    \caption{Same as Fig.~\ref{fig:t_vs_freqs}, but displaying the post-fit timing residuals (in $\mu$s) for the y-axis.
    }
    \includegraphics[width=\linewidth-.5cm, height=22cm]{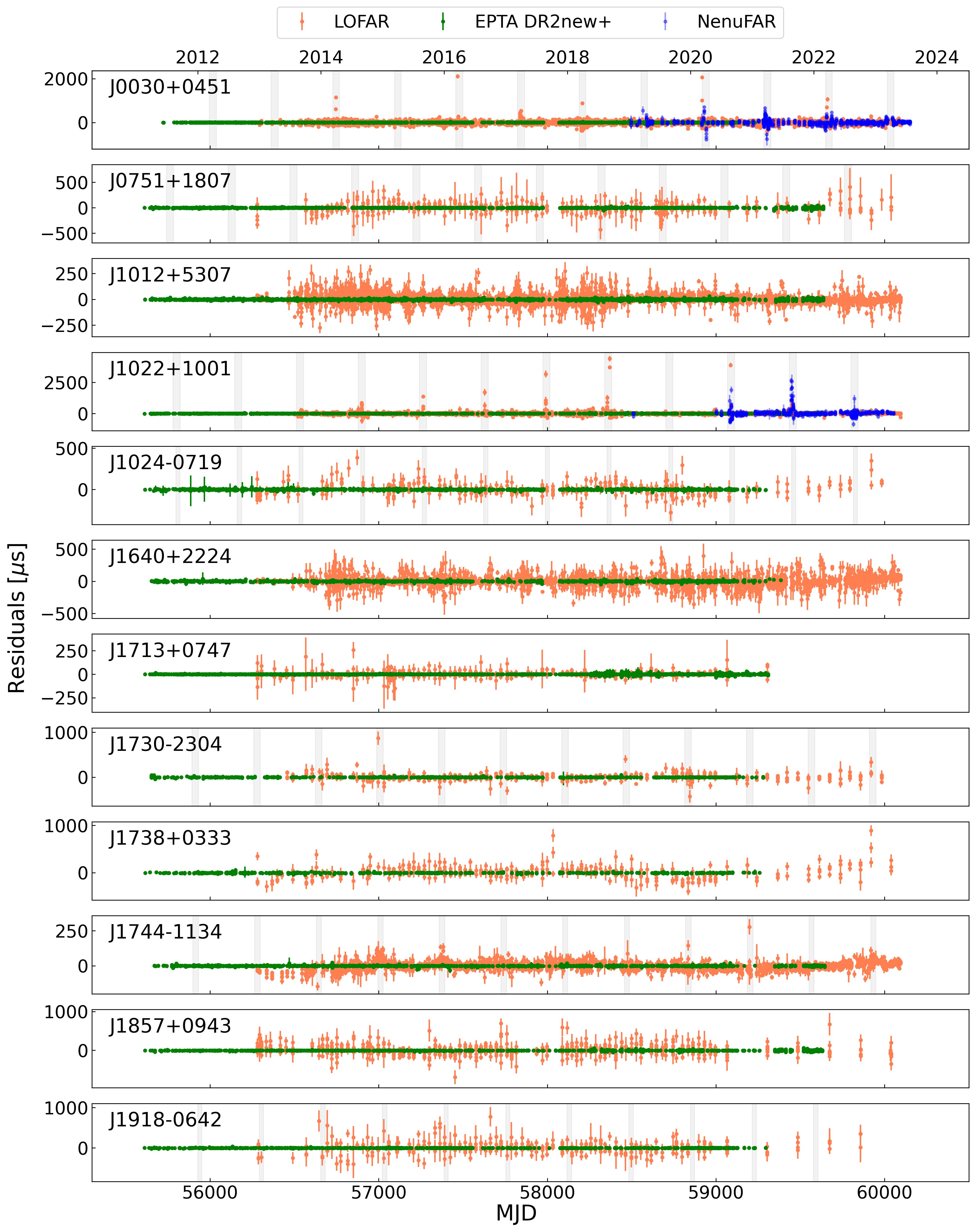}
    \label{fig:t_vs_res}
\end{figure*}
%\FloatBarrier
\iffalse
\begin{figure}[ht!]
    \centering
    \caption{Same as Fig.~\ref{fig:t_vs_freqs}, but displaying the post-fit timing residuals (in $\mu$s) for the y-axis.}
    \includegraphics[width=\linewidth-.5cm]{Figures/Data/EPTA_LOFAR_residuals.jpg}
    \label{fig:t_vs_res}
\end{figure}
\fi
\end{appendix}

\end{document}